\RequirePackage{tikz}
\documentclass[pdflatex,sn-mathphys]{sn-jnl}

\jyear{2021}%

\theoremstyle{thmstyleone}%
\usepackage{amsmath,amsfonts}
\usepackage{bm}
\usepackage{graphicx}

\usepackage{physics}
\usepackage{caption}
\usepackage{subcaption}

\usepackage{tikz}
\usetikzlibrary{shapes.misc}
\usetikzlibrary{calc}
\usetikzlibrary{arrows}
\usetikzlibrary{shapes}
\usetikzlibrary{decorations.pathreplacing,decorations.markings,decorations.text}

\definecolor{reddark}{RGB}{179,0,59}
\definecolor{greendark}{RGB}{0,225,0}
\definecolor{greendark2}{RGB}{0,200,0}
\definecolor{purple2}{RGB}{147, 81, 22}
\definecolor{grey2}{RGB}{144, 148, 151}
\definecolor{yellow2}{RGB}{255, 216, 1}
\usepackage{xfp}

\tikzset{cross/.style={cross out, draw=black, minimum size=2*(#1-\pgflinewidth),very thick, inner sep=0pt, outer sep=0pt}, cross/.default={1pt}}

\def\vv{{\bm{v}}}
\DeclareMathAlphabet{\mathsfit}{\encodingdefault}{\sfdefault}{m}{sl}
\SetMathAlphabet{\mathsfit}{bold}{\encodingdefault}{\sfdefault}{bx}{n}
\newcommand{\tens}[1]{\bm{\mathsfit{#1}}}
\def\tT{{\tens{T}}}
\def\tZ{{\tens{Z}}}
\def\gG{{\mathcal{G}}}
\def\evv{{v}}
\def\emM{{M}}
\def\etT{{T}}
\def\mM{{\bm{M}}}

\theoremstyle{thmstyletwo}%

\theoremstyle{thmstylethree}%

\begin{document}

\title[]{The Tensor Track VII:
From Quantum Gravity to Artificial Intelligence}

\author*[1]{\fnm{Mohamed} \sur{Ouerfelli}}\email{Mohamed-oumar.ouerfelli@cea.fr}
\author[2]{\fnm{Vincent} \sur{Rivasseau}}\email{Vincent.rivasseau@ijclab.in2p3.fr}
\equalcont{These authors contributed equally to this work.}
\author[1]{\fnm{Mohamed} \sur{Tamaazousti}}\email{Mohamed.tamaazousti@cea.fr}
\equalcont{These authors contributed equally to this work.}

\affil*[1]{Universit\'e Paris-Saclay, CEA-LIST, F-91120 Palaiseau, France}

\affil[2]{Universit\'e Paris-Saclay, CNRS/IN2P3, IJCLab, 91405 Orsay, France}

\abstract{Assuming some familiarity 
with 
quantum field theory
and
with the tensor track approach 
that one of us presented in the previous series “Tensor Track I-VI”,
we provide, as usual, the developments in quantum gravity
of the last two years. 
Next we  present in some  detail two algorithms inspired by Random Tensor Theory which has been developed in the quantum gravity context. One is devoted to the detection and recovery of of a signal
in a random tensor (that can be associated to the noise) with new theoretical guarantees for more general cases such as tensors with different dimensions. The other, SMPI, is more ambitious but maybe less rigorous; it is devoted to significantly and fundamentally improve the performance of algorithms for Tensor principal component analysis but without complete theoretical guarantees yet.
Then we sketch all sorts of application 
relevant to information theory and artificial intelligence and provide their corresponding bibliography.}




\maketitle

\section{Introduction}\label{sec1}

Random tensors originated in theoretical physics.  Matrix models had 
 some success
 in quantizing the strong interaction, as
t'Hooft made the fundamental observation \cite{t1993planar} that their $1/N$ expansion is a topological expansion.

In 1990's the dominating theory was string theory.
Random  matrix  models were seen at this  time as a successful theory for quantizing gravity but only in two dimensions. 
The inventors of random tensor models wanted to replicate the success of matrix models for dimensions $D=3,4...$ \cite{ambjorn1991three,gross1992tensor,sasakura1991tensor}.

In 2010 a new kind of 1/N expansion was discovered for these random tensor models (\cite{gurau20111}, for a review, consult \cite{guruau2017random}). It relies heavily on the Gurau degree of a $D+1$-colored graph $G$, with is a concept partly topological and partly combinatorial.
The Gurau degree $\omega$ is a positive number $\omega (G) \in {\mathbb N}$.
To define it, we need a new notion, that of the jackets. For the moment, it suffices to say that a jacket $J_{\sigma}$ is associated 
to a colored graph and a cyclic permutation $\sigma$ on the colors.
There are three  inequivalent jackets for $D=3$ \cite{gurau20111}, twelve  inequivalent jackets for $D=4$ and so on \cite{gurau20111a}.
To each jacket is associated a combinatorial map, hence a matrix model, thus a degree of the ordinary $1/N$ graphs of the corresponding matrix model.
The Gurau degree is proportional to the {\it sum} of the ordinary (t'Hooft) degree of the jackets :

\begin{equation} \omega (G) = \frac{1}{(D-1)!}   \sum_{\sigma} g(J_{\sigma} )
\end{equation} 

Initially, the tensor track 
 is an attempt to quantize gravity in dimension $D>2$  by combining random tensor models to discrete geometry and the  renormalisation  group  \cite{rivasseau2012quantum}. 
The tensor track lies at the crossroad of  several  closely  related  approaches to quantize gravity,  most  notably  
causal dynamical  triangulations \cite{loll2019quantum},  quantum field theory on non-noncommutative spaces \cite{szabo2003quantum,grosse2014self,branahl2021scalar}, and group field theory \cite{oriti2016group}. 
   
Random tensors share with random matrices the fact that they are a zero-dimensional world, and, as such, they are background-independent; it makes no references whatsoever of any particular space-time. Moreover,  random  tensors models, based on the field theory of Feynman, are amenable  to renormalisation group techniques \cite{geloun2013renormalizable}. Simple models even share with non-Abelian gauge theories the  property of asymptotic freedom \cite{ben20133d,rivasseau2015tensor}.  

Random tensors are expected to play a growing role in many areas of mathematics, physics, and computer science. To our knowledge, 
communities using random tensors have mostly grown apart, developing their own tools and results.
Nowadays there is an increasing circle of mathematicians and physicists working on random tensors, and these people have been lately inclined towards more applications linked to data analysis and artificial intelligence. In this way, the line which separated them from computer scientists becomes a little bit blurred. 

This review is organized as follows. Section II is devoted to a brief summary of   the tensor track. 
In section III we review in more depth two tensors-inspired algorithms.  In Section IV we made the rather bold step
of applying this stuff to artificial intelligence. Section V presents our future perspectives.

\section{Quantum gravity}\label{sec2}

Let us review briefly the previous chapters of the tensor  track \cite{rivasseau2012quantum,rivasseau2014tensor,delporte2018tensor,delporte2020tensor}.

In the Hermitian matrix ensemble (GUE) perturbed by a quadratic interaction,  the $1/N$ expansion is well known. 
The  free partition function is $ \int d M e^{- \frac{N}{2}\Tr M^2 } $, where 
\begin{equation}
\quad d M= \prod_k d M_{kk}. \prod_{i<j} d \Re M_{ij}d \Im M_{ij},
\end{equation}
and the expectation values of $U(N)$ invariants 
\begin{equation}
< \Tr M^{p_1} \Tr M^{p_2}...\Tr M^{p_k}
 >
\end{equation}
 is entirely determined by the propagator 
\begin{equation} C_{ij,kl} =  \frac{1}{N} \delta_{il}\delta_{jk}
\end{equation}
and by Wick's rule.
 
Any scalar function of a tensorial quantum field theory  
is a big functional integral
 on a Gaussian measure and an interactive part. 
In the tensorial case
this interactive part is a sum of invariants of the tensor.
For example the partition function is a scalar function of $N$, defined by
\begin{equation}
Z(N) =  
\int d \mu (T) e^{-
 \sum_{Inv} S_{Inv}( T) }
\end{equation}
so as 
the free energy. The partition function and the corresponding free energy are related by a normalized  logarithm 
\begin{equation}
F(N) =\frac{1}{N^D} \log Z(N) 
\end{equation}

\begin{figure}[htb]
  \centering
 \begin{subfigure}{.16\textwidth}
 \centering
\vspace{0.4cm}
\begin{tikzpicture}[>=latex,scale=0.15]
    \begin{scope}
    \coordinate (A) at (-4.142,-10) ;
    \coordinate (B) at (4.142,-10) ;
    \draw[very thick, red,-] (A) -- (B)  ;
    \draw[very thick, greendark,-] (A) arc (-90-57/2:-90+57/2:8.69)   ;
    \draw[very thick, blue,-] (A) arc (90+57/2:90-57/2:8.69)   ;
    \node[draw,circle,inner sep=1.75pt,fill,black] at (A){};
    \node[draw,circle,inner sep=1.75pt,fill,white] at (B){};
    \node[draw,circle,inner sep=1.75pt,black] at (B){};
    \end{scope}
\end{tikzpicture}
\vspace{0.3cm}
  \caption{Melon
  graph}
  \label{melon3}
 \end{subfigure} 
 \hspace{1cm} 
\begin{subfigure}{.2\textwidth}
 \centering
\begin{tikzpicture}[>=latex,scale=0.2]
    \begin{scope}
    \coordinate (A) at (-3.142,-3.142) ;
    \coordinate (B) at (3.142,-3.142) ;  
    \coordinate (C) at (3.142,3.142) ;
    \coordinate (D) at (-3.142,3.142) ;
    \draw[very thick, red,-] (A) -- (B)  ;
    \draw[very thick, red,-] (C) -- (D)  ;
    \draw[very thick, blue,-] (A) -- (D)  ;
    \draw[very thick, blue,-] (C) -- (B)  ;
    
    \draw[very thick, greendark,-] (D) arc (-90-57/2:-90+57/2:7) ;
    \draw[very thick, greendark,-] (A) arc (90+57/2:90-57/2:7) ;
    \node[draw,circle,inner sep=1.75pt,fill,black] at (A){};
    \node[draw,circle,inner sep=1.75pt,fill,white] at (B){};
    \node[draw,circle,inner sep=1.75pt,black] at (B){};
    \node[draw,circle,inner sep=1.75pt,fill,black] at (C){};
    \node[draw,circle,inner sep=1.75pt,fill,white] at (D){};
    \node[draw,circle,inner sep=1.75pt,black] at (D){};
    \end{scope}
\end{tikzpicture}
  \caption{Pillow graph}
  \label{Pillow3}
 \end{subfigure}
  \hspace{1cm} 
\begin{subfigure}{.2\textwidth}
 \centering
\begin{tikzpicture}[>=latex,scale=0.2]
    \begin{scope}
    \coordinate (A) at (-3.542, 2.142) ;
    \coordinate (B) at (0,4) ;
    \coordinate (C) at (3.542,2.142) ;
    \coordinate (D) at (3.542,-2.142) ;
    \coordinate (E) at (0,-4) ;
    \coordinate (F) at (-3.542,-2.142) ;
    
    \draw[very thick, red,-] (A) -- (B)  ;
    \draw[very thick, red,-] (C) -- (D)  ;
    \draw[very thick, red,-] (E) -- (F)  ;
    \draw[very thick, blue,-] (B) -- (C)  ;
    \draw[very thick, blue,-] (D) -- (E)  ;
     \draw[very thick, blue,-] (F) -- (A)  ;
    \draw[very thick, greendark,-] (A) -- (D) ;
    \draw[very thick, greendark,-] (B) -- (E) ;
    \draw[very thick, greendark,-] (C) -- (F) ;
    \node[draw,circle,inner sep=1.75pt,fill,black] at (A){};
    \node[draw,circle,inner sep=1.75pt,fill,white] at (B){};
    \node[draw,circle,inner sep=1.75pt,black] at (B){};
    \node[draw,circle,inner sep=1.75pt,fill,black] at (C){};
    \node[draw,circle,inner sep=1.75pt,fill,white] at (D){};
    \node[draw,circle,inner sep=1.75pt,black] at (D){};
    \node[draw,circle,inner sep=1.75pt,fill,black] at (E){};
    \node[draw,circle,inner sep=1.75pt,fill,white] at (F){};
    \node[draw,circle,inner sep=1.75pt,black] at (F){};
    \end{scope}
\end{tikzpicture}
  \caption{$K_{3,3}$ graph}
  \label{Tetra3}
 \end{subfigure}
  \caption{Examples of $U(N)$ invariants}
  \label{coninv1}
\end{figure}
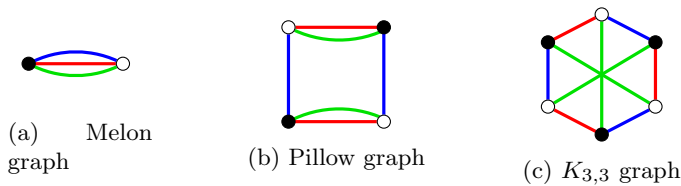

The invariants themselves can be classified in terms of graphs. Of course these graphs depend upon the group symmetries of the tensor. 
For matrix models 
the expectation values of the invariants can be classified by ribbon diagrams.
In the case of tensor models  Figure \ref{coninv1} depicts a partial list of connected invariants for $\bigotimes_{i=1}^3 U(N)$.

To generalize the $1/N$ expansion to the tensor case, the first step is to choose an invariant, for example a quartic invariant, and 
to normalize the $S_{Inv_0}(T)$ for {\it that} invariant:
\begin{equation}
S_{Inv_0}( T) \to  \frac{\lambda}{N^\alpha} S_{Inv_0}( T) 
\end{equation}
Now the partition function and the corresponding free energy depend on {\it  two variables}. In a quantum field theory the usual form of perturbation theory is to expand in power series in $\lambda$, and the perturbation is indexed by Feynman amplitudes associated to Feynman graphs.
For instance the perturbation of the free energy is of the following form  
\begin{equation}
F(\lambda,N)= \sum_{G} \frac{(-\lambda)^{v(G)} }{sym(G)}A ( G,N) 
\end{equation}

Once this is done, 
the hard step is to find $\alpha$ such that 1/N expansion exists, i.e.
such that the perturbation of the free energy is of the following form  
\begin{equation}
F(\lambda,N)= \sum_{G} \frac{(-\lambda)^{v(G)} }{sym(G)}A ( G,N) = \sum_{\omega  \in {\mathbb N} } N^{- \omega} F_{\omega} (\lambda ) 
\end{equation}

In the case of  $\bigotimes_{i=1}^D U(N)$ and a quartic $D$-melonic invariant,  the $1/N$ expansion is governed by the Gurau degree  and 
$F_{0} (\lambda ) $ is formed by all the $D+1$ melonic graphs (the 0 color being  associated to the Feynman propagators \cite{bonzom2012random}).
The family of $D+1$ melonic graphs \cite{bonzom2011critical} which lead  the 1/N expansion 
of random tensors models can perhaps be called too trivial from a topologist point of view; it corresponds to some triangulations of the sphere $S_D$.
But, as the Gurau degree is not only purely topological, the interplay between combinatorics and topology in   
sub-leading terms can be amazingly complex, even for a geometrically-oriented person!

Now that we have been able to identify the leading terms in the $1/N$ expansion, the second step is to be able to resum them, i.e. to explicitly 
compute $F_{0} (\lambda ) $. This step has been performed for the first time by a paper by Gurau and Ryan \cite{gurau2014melons}.
 From a probabilistic and statistical mechanical point of view, it corresponds to Aldous phase of branched polymers. 

Once we have been able to compute explicitly $F_{0} (\lambda ) $,
many possibilities are open to us:
 modifying the symmetry of the main tensor,  include the renormalization group by modifying 
in a specific way our propagators, a non-perturbative treatment of some simple models...

From the perspective of matrix models, to go further in the two parameters approach require a particular technique, namely double scaling. 
The first step of applying this technique to tensors has been done \cite{gurau2016regular,dartois2013double}.
 The initial papers have been followed by mixed results, some results suggest the universality of branched polymers, 
 others pointing to the fact that some 
simple and natural restrictions change that universality class.
But, from the perspective of quantum gravity, the main goal is to resum the sub-leading terms
in order to find a more interesting phase of geometry pondered by Einstein-Hilbert action.
In this vein, we would like to highlight one contribution, that of Lionni and Marckert \cite{lionni2021iterated}. In this paper they use new combinatorial 
bijections to uncover 
a random phase in higher dimensions.

Let us come to the Sachdev-Ye-Kitaev (SYK) model \cite{maldacena2016remarks,gross2017generalization}. Discovered by Kitaev, it is a quartic model of N Majorana fermions coupled by a disordered tensor. It is a model of condensed matter, hence it depends on time though a Hamiltonian.
The disordered tensor is centered Gaussian iid
\begin{equation}
< J_{abcd} > = 0 ,  \quad < J^2_{abcd} >= \frac{\lambda^2}{N^3},
\end{equation}
and the Hamiltonian  is simply $H = J_{abcd} \psi_a \psi_b \psi_c \psi_d$.
This model posses three important properties: it is solvable at large $N$, there is a conformal symmetry at strong coupling, hence it can be a fixed point of the renormalization group,
 and, above all from quantum gravity, it is maximally chaotic in the sense of \cite{maldacena2016bound}. Hence the SYK model, although very simple,
 offers a path to the main theoretical concepts of quantum gravity,
 such as Bekenstein-Hawking entropy and holography. 

SYK became a very active field, from the early papers to nowadays. At large $N$
the Schwinger-Dyson equation for the 2-point function is closed. The conformal  symmetry 
can be broken and  the corresponding subject goes under the name of
near-$AdS_2$/near-$CFT_1$ correspondence. 
This  entails a relationship with Jackiw-Teitelboim two-dimensional  quantum gravity.

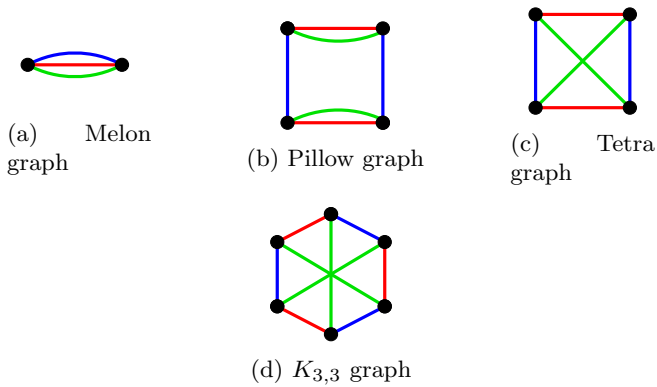
\begin{figure}[htb]
  \centering
 \begin{subfigure}{.16\textwidth}
 \centering
\vspace{0.4cm}
\begin{tikzpicture}[>=latex,scale=0.15]
    \begin{scope}
    \coordinate (A) at (-4.142,-10) ;
    \coordinate (B) at (4.142,-10) ;
    \draw[very thick, red,-] (A) -- (B)  ;
    \draw[very thick, greendark,-] (A) arc (-90-57/2:-90+57/2:8.69)   ;
    \draw[very thick, blue,-] (A) arc (90+57/2:90-57/2:8.69)   ;
    \node[draw,circle,inner sep=1.75pt,fill,black] at (A){};
    \node[draw,circle,inner sep=1.75pt,fill,black] at (B){};
    \end{scope}
\end{tikzpicture}
\vspace{0.3cm}
  \caption{Melon graph
  }
  \label{melon}
 \end{subfigure}  
  \hspace{1cm}
\begin{subfigure}{.2\textwidth}
 \centering
\begin{tikzpicture}[>=latex,scale=0.2]
    \begin{scope}
    \coordinate (A) at (-3.142,-3.142) ;
    \coordinate (B) at (3.142,-3.142) ;  
    \coordinate (C) at (3.142,3.142) ;
    \coordinate (D) at (-3.142,3.142) ;
    \draw[very thick, red,-] (A) -- (B)  ;
    \draw[very thick, red,-] (C) -- (D)  ;
    \draw[very thick, blue,-] (A) -- (D)  ;
    \draw[very thick, blue,-] (C) -- (B)  ;
    
    \draw[very thick, greendark,-] (D) arc (-90-57/2:-90+57/2:7) ;
    \draw[very thick, greendark,-] (A) arc (90+57/2:90-57/2:7) ;
    \node[draw,circle,inner sep=1.75pt,fill,black] at (A){};
   \node[draw,circle,inner sep=1.75pt,fill,black] at (B){};
    \node[draw,circle,inner sep=1.75pt,black] at (B){};
    \node[draw,circle,inner sep=1.75pt,fill,black] at (C){};
    \node[draw,circle,inner sep=1.75pt,fill,black] at (D){};
    \node[draw,circle,inner sep=1.75pt,black] at (D){};
    \end{scope}
\end{tikzpicture}
  \caption{Pillow graph}
  \label{Pillow2}
 \end{subfigure}
  \hspace{1cm} 
  \begin{subfigure}{.16\textwidth}
 \centering
\begin{tikzpicture}[>=latex,scale=0.15]
    \begin{scope}
    \coordinate (A) at (-4.142,-4.142) ;
    \coordinate (B) at (4.142,-4.142) ;
    \coordinate (C) at (4.142,4.142) ;
    \coordinate (D) at (-4.142,4.142) ;
    \draw[very thick, red,-] (A) -- (B)  ;
    \draw[very thick, red,-] (C) -- (D)  ;
    \draw[very thick, blue,-] (A) -- (D)  ;
    \draw[very thick, blue,-] (C) -- (B)  ;
    \draw[very thick, greendark,-] (A) -- (C)  ;
    \draw[very thick, greendark,-] (D) -- (B)  ;
    \node[draw,circle,inner sep=1.75pt,fill,black] at (A){};
    \node[draw,circle,inner sep=1.75pt,fill,black] at (B){};
    \node[draw,circle,inner sep=1.75pt,fill,black] at (C){};
    \node[draw,circle,inner sep=1.75pt,fill,black] at (D){};
    \end{scope}
\end{tikzpicture}
  \caption{Tetra graph 
  }
  \label{Tetra}
 \end{subfigure}
   \hspace{1cm} 
\begin{subfigure}{.2\textwidth}
 \centering
\begin{tikzpicture}[>=latex,scale=0.2]
    \begin{scope}
    \coordinate (A) at (-3.542, 2.142) ;
    \coordinate (B) at (0,4) ;
    \coordinate (C) at (3.542,2.142) ;
    \coordinate (D) at (3.542,-2.142) ;
    \coordinate (E) at (0,-4) ;
    \coordinate (F) at (-3.542,-2.142) ;
    
    \draw[very thick, red,-] (A) -- (B)  ;
    \draw[very thick, red,-] (C) -- (D)  ;
    \draw[very thick, red,-] (E) -- (F)  ;
    \draw[very thick, blue,-] (B) -- (C)  ;
    \draw[very thick, blue,-] (D) -- (E)  ;
     \draw[very thick, blue,-] (F) -- (A)  ;
    \draw[very thick, greendark,-] (A) -- (D) ;
    \draw[very thick, greendark,-] (B) -- (E) ;
    \draw[very thick, greendark,-] (C) -- (F) ;
    
    \node[draw,circle,inner sep=1.75pt,fill,black] at (A){};
    \node[draw,circle,inner sep=1.75pt,fill,black] at (B){};
    \node[draw,circle,inner sep=1.75pt,black] at (B){};
    \node[draw,circle,inner sep=1.75pt,fill,black] at (C){};
    \node[draw,circle,inner sep=1.75pt,fill,black] at (D){};
    \node[draw,circle,inner sep=1.75pt,black] at (D){};
    \node[draw,circle,inner sep=1.75pt,fill,black] at (E){};
    \node[draw,circle,inner sep=1.75pt,fill,black] at (F){};
    \node[draw,circle,inner sep=1.75pt,black] at (F){};
    \end{scope}
\end{tikzpicture}
  \caption{$K_{3,3}$
  graph}
 \end{subfigure}
  \caption{Examples of $O(N)$} graphs and their associated invariants 
  \label{coninv2}
\end{figure}

Witten has found a genuine field theory model (with no disorder),  in which the tensors plays a much more fundamental role \cite{witten2019syk}. 
In a nutshell, he discovered   that his model has the same melonic limit as the tensors models pioneered by Gurau.
Klebanov and Tarnopolsky \cite{klebanov2017uncolored}, when combined with an earlier work of Carrozza and Tanasa \cite{carrozza2016n},
allows on a big simplification of the group symmetry of the main tensor, from $U(N)^{D(D-1)/2}$ to $O(N)^{D}$ (see Figure \ref{coninv2}).

The action of the KTCT model is
\begin{equation}
S= \int dt \frac{i}{2} \psi_{i_1,i_2,i_3} \partial_t \psi_{i_1,i_2,i_3}+\frac{\lambda}{4N^{3/2}}  \psi_{i_1,i_2,i_3}  \psi_{i_4,i_5,i_3}  \psi_{i_4,i_2,i_6}  \psi_{i_1,i_5,i_6} .
\end{equation}

Unlike the initial SYK model, these tensor models  fit in the framework of local quantum field theory with $D=1$.
Hence there is a possibility to extend them
in $D>1$!
We would like to stressed one recent contributions in this domain.
For $p = 3 $ and $p=5$ there exist a melonic large $N$ limit for $p$-irreducible tensors in the sense of Young tableaux \cite{benedetti20191,carrozza2019syk,carrozza2022melonic}. 
Carrozza and Harribey overcome huge difficulties to solve the case $p=5$ \cite{carrozza2022melonic}.

\section{Algorithms}

Given the richness of the tools developed in the subject of random tensors, it is very appealing to attempt to take advantage of them for various other situations outside of quantum gravity.

In this section, we consider an important problem, Tensor PCA, where these tools were successfully used to introduce a novel framework that achieved new theoretical and practical improvements. Tensor PCA  \cite{richard2014statistical} consists in detecting and retrieving a spike $\vv_0^{\otimes k}$ from noise-corrupted multi-linear measurements in the form of a tensor $\tT$
\begin{equation}
    \tT=\beta \vv_0^{\otimes k}+\tZ,
\end{equation}
with $\tZ \in (\mathbb{R}^n)^{\otimes k}$ a pure Gaussian noise tensor of order k and dimension $n$ with identically independent distributed (iid) standard Gaussian entries: $\tZ_{i_1,i_2,\dots,i_k} \sim \mathcal{N}(0,1)$ and $\beta$ is the signal-to-noise ratio. 

\subsection{Motivation}
The motivation for Tensor PCA is two-fold:
\begin{itemize}
\item Practical applications : algorithms addressing Tensor PCA could easily be generalized to Tensor decomposition, which have applications in Tensor faces \cite{vasilescu2002multilinear}, Hyperspectral imagery \cite{nasrabadi2013hyperspectral}, DNN compression \cite{astrid2017cp, wang2020cpac}, multimodal data fusion  \cite{lahat2015multimodal}, wireless communication \citep{decurninge2020tensor}, computer vision \citep{panagakis2021tensor}, natural language processing  \citep{sobhani2019text}, etc. More details could be found in Section 4.

\item  Theoretical interest: Tensor PCA  is often used as a prototypical inference problem for the theoretical study of the computational hardness of optimization in high-dimensional non-convex landscapes, in particular using the well spread gradient descent algorithm and its variants (\cite{arous2020algorithmic, mannelli2019passed,mannelli2019afraid,mannelli2020marvels}). Tensor PCA is also considered as an interesting study case of statistical-algorithmic gaps that appears in various other problems (see references in \citep{arous2020algorithmic} and \citep{luo2020open}). Indeed, while information theory shows that is theoretically possible to recover the signal for $\beta \sim O(1)$, all existent algorithms have been shown or conjectured to have an algorithmic threshold of at least $\beta \sim O(n^{(k-2)/4})$. 
\end{itemize}


\subsection{Random Tensor Theory for Tensor PCA}
\subsubsection{The framework}
Matrix data analysis and principal component analysis (PCA) is mostly stated in the  ``quantum-mechanics" language of eigenvalues rather than in the ``quantum-field theoretic" language of  invariants and (Feynman) graphs. For  tensors the quantum-field language is the natural one. An important task in tensor data analysis is therefore to {\bf translate}
the results of matrix data analysis and PCA into the  quantum-field theoretic language
of  invariants and graphs.


An original connection have been made in \cite{Ouerfelli2022random} between tensorial data analysis and the random tensor theory developed last decade in the context of quantum gravity by the theoretical physics community.

\begin{figure}[h!]
    \centering
    \includegraphics[width=0.7\textwidth]{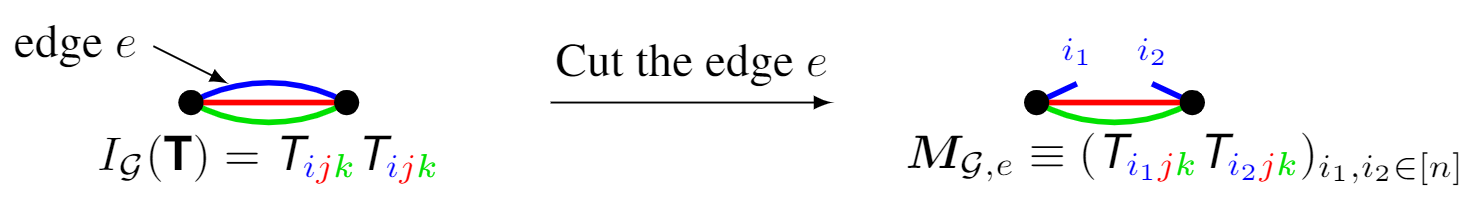}
    \caption{Cutting an edge}
    \label{fig:FG}
\end{figure}
This connection is based on the introduction of matrices that are built out of a graph and an edge as illustrated in Figure \ref{fig:FG}. Indeed, given a graph invariant, we call "cutting an edge" the fact of not performing a sum over the index associated to this edge, which gives us a matrix.

The eigenvector associated to the largest eigenvalue of this matrix can then be proven to be correlated to the signal vector $\vv$ for a range of SNR $\beta$. The algorithm \ref{algo:recovery} is based on this simple fact.

\begin{algorithm}[h!]
  \caption{Recovery algorithm associated to the graph $\gG$ and edge $e$}
    \label{algo:recovery}
\begin{algorithmic}
  \State {\bfseries Input:} The tensor $\tT=\beta \vv^{\otimes k} + \tZ $
  \State {\bfseries Goal:} Estimate $\vv_0$.
  \State Calculate the matrix $\mM_{\gG,e} (\tT)$ 
  \State Compute its top eigenvector by matrix power iteration (repeat $\evv_i \leftarrow \emM_{ij} \evv_j$).
  \State {\bfseries Output:} Obtaining an estimated vector $\vv$
\end{algorithmic}
\end{algorithm}

\subsubsection{The algorithms}

\begin{figure}[h!]
    \centering
    \includegraphics[width=0.6\textwidth]{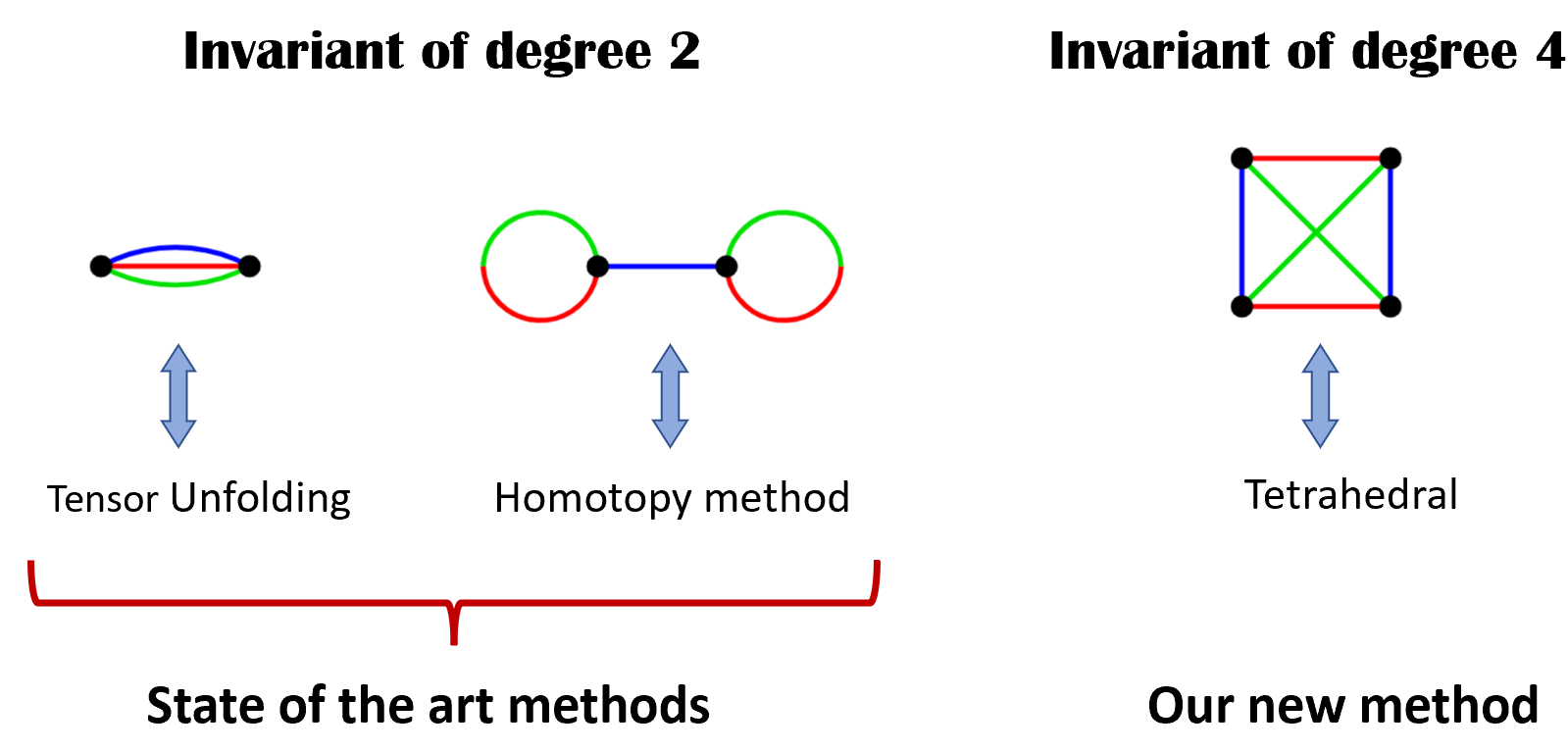}
    \caption{Methods associated to invariant graphs}
    \label{fig:SOTA}
\end{figure}

It appears that the two state of the art (SOTA) methods, which are the Tensor Unfolding method and the Homotopy-inspired method, are equivalent to the algorithms associated to the graphs of degree 2. This striking fact incites us to investigate the algorithm associated to the tetrahedral graph which is a graph of degree 4 as illustrated in Figure \ref{fig:SOTA}.

{\bf Novel theoretical threshold:}
Let's first consider the more general case where the tensor $\tT$ has axes of different dimensions $n_i$ ($\tT \in \bigotimes_{i=1}^k \mathbb{R}^{n_i}$). We can assume without any loss of generality that $n_1 \geq n_2 \geq \dots \geq n_k$. 
\begin{equation}
\displaystyle
    \tT= \beta \vv_1\otimes \dots \otimes \vv_k+\tZ  \;\;\;\; \text{where} \;\;\;\;  \vv_i \in \mathbb{R}^{n_i},  \;\;\;\; n_i \in \mathbb{N}.
\end{equation}

Our framework allows us to derive a new algorithmic threshold for this more general case : the threshold for $\vv_1$ is given by $\max \left( (\prod_{i=1}^k n_i)^{1/4}, n_1^{1/2} \right)$ while the thresholds for $\vv_{j}, \; j\geq 2$ are equal to $(\prod_{i=1}^k n_i)^{1/4}$. It is, to the best of our knowledge, the first generalization of the threshold $\beta=n^{k/4}$ derived in \cite{richard2014statistical} when $n_i=n \; \forall i \in [k]$. 

\begin{figure}
    \centering
    \begin{subfigure}{.5\textwidth}
    \includegraphics[width=0.95\textwidth]{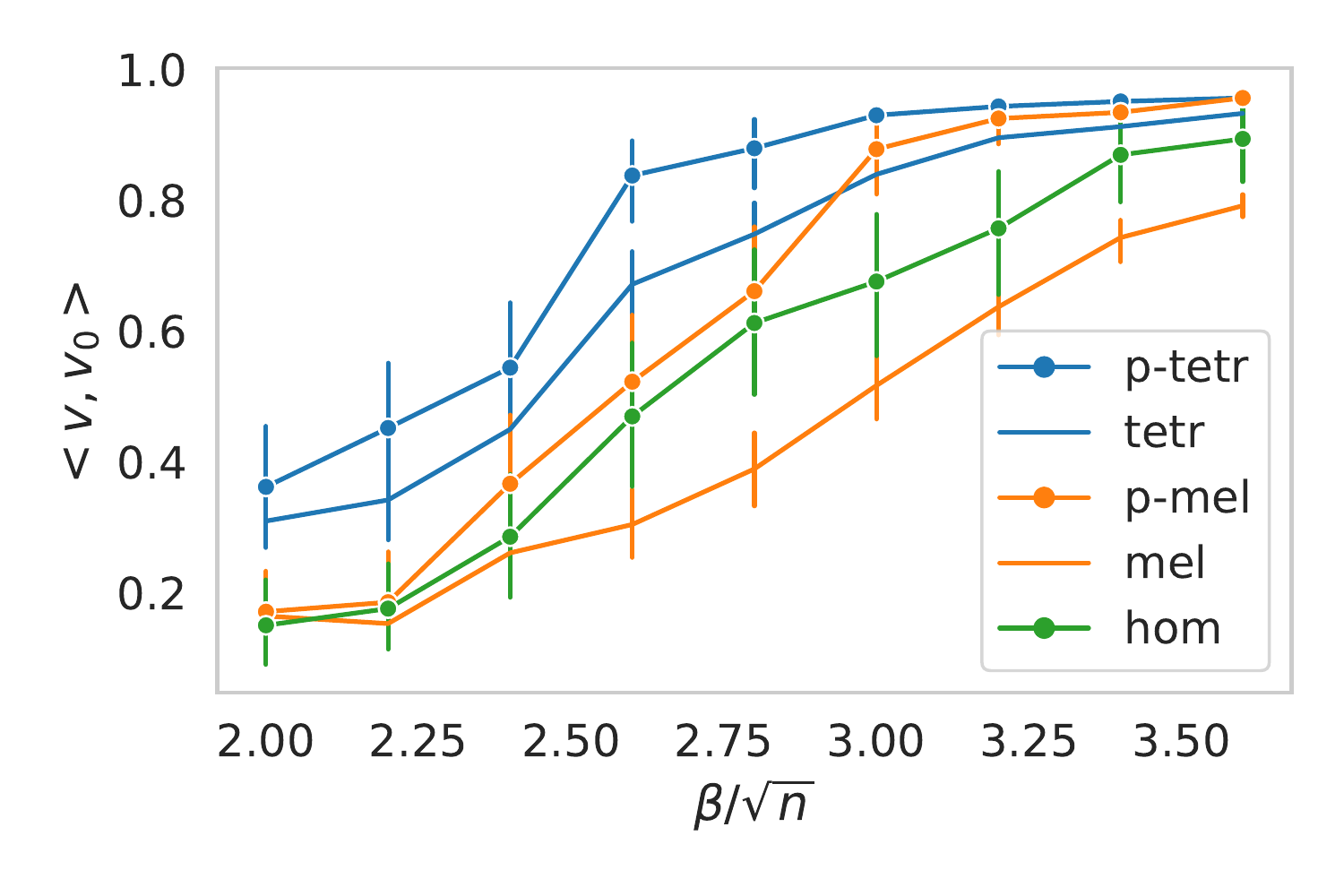}
    \end{subfigure}%
    \begin{subfigure}{.5\textwidth}
    \centering
    \includegraphics[width=0.95\textwidth]{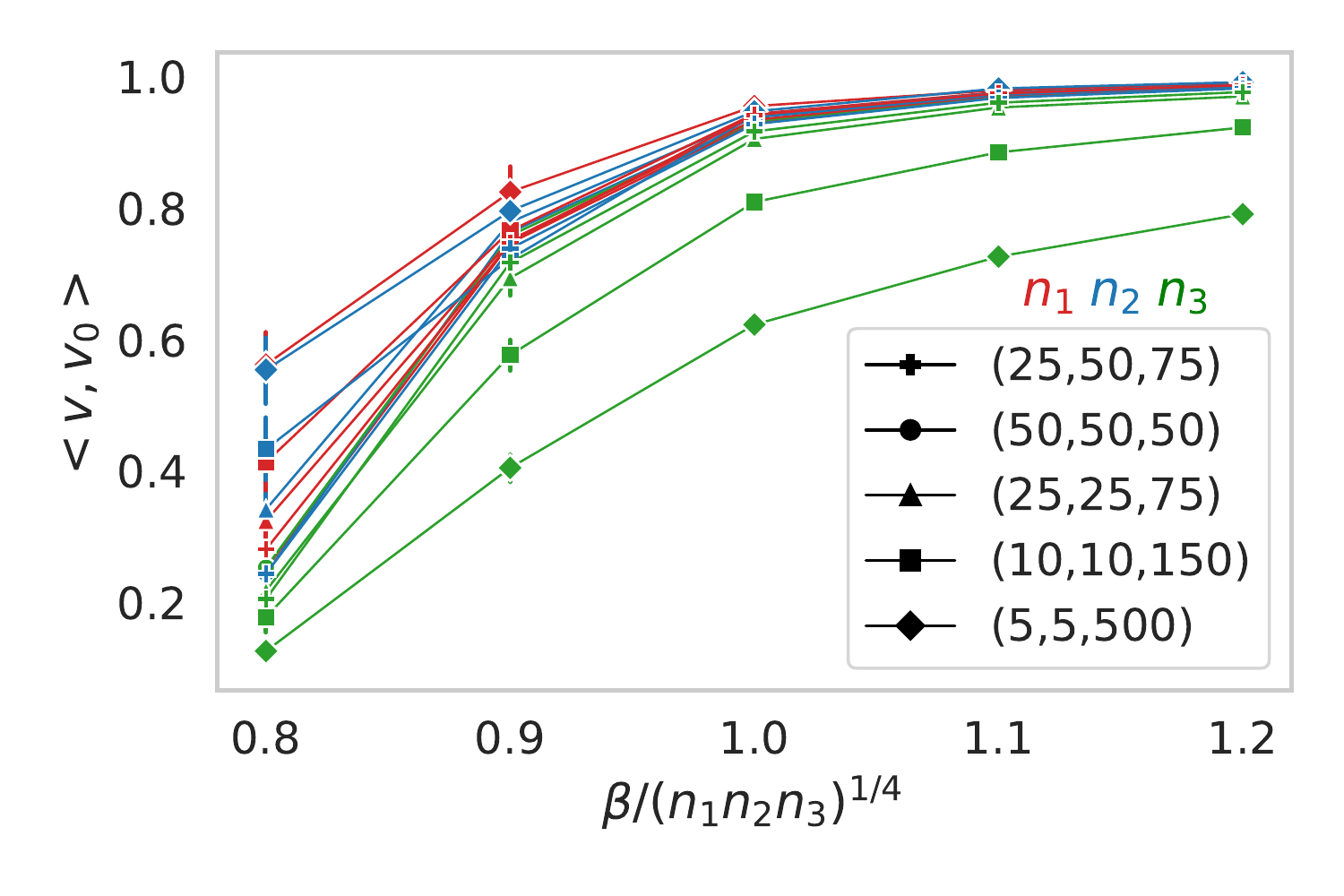}
    \end{subfigure}
    \caption{In the right : (a) Comparison of different methods for symmetric
recovery. n=150. In the left : (b)  Recovery of a spike with different dimensions}
    \label{fig:Comp}
\end{figure}

{\bf Tensor PCA experiments:} In Figure \ref{fig:Comp} right, we compare the recovery performance. For every algorithm we use two variants: the simple algorithm outputting $\vv$ and an algorithm where we apply $100$ power iterations on $\vv$: $\evv_i \leftarrow \etT_{ijk} \evv_{j} \evv_{k}$, distinguishable by a prefix "p-". We compare our method (tetrahedral) to other algorithmic methods: the melonic (tensor unfolding) and the homotopy. They give the state of art respectively for the symmetric and asymmetric tensor. We see that our new algorithm is able to achieve better performance. However, it should be noted that the tetrahedral method is more costly in computational power given that it is based on a graph of degree 4. In conclusion, this novel framework is not only able to cover the two SOTA methods as special case of algorithms, but it is also able to offer novel algorithms with larger computational cost but better performance. Providing a large variety of algorithms with different performance/computational cost ratio is a very interesting feature as it is adaptable to the resources and precision objective of the situation.

We see in the Figure \ref{fig:Comp} left, that the threshold $(n_1 n_2 n_3)^{1/4}$ for the three vectors matches perfectly with the experiences when $n_3 < n_1 . n_2$. We also see that when $n_3 > n_1 . n_2$ the recovery of $v_3$ (in green and with the diamond and square markers) have a different asymptotic behavior than $v_1$ and $v_2$ (it becomes $n_3^{1/2}$ since $n_3^{1/2}\geq (n_1 n_2 n_3)^{1/4}$), corresponding to our theoretical predictions. 


\subsection{Selective Multiple Power Iteration for high dimensional non convex landscapes.}

The previous framework is able to provide algorithms with theoretical guarantees that are easily obtained thank to Random Tensor Theory and that can address more general situations like a tensor with different dimensions.

In contrast, the second algorithm SMPI \cite{ouerfelli2021selective} (Selective Multiple Power Iteration) does not have theoretical guarantees yet, but it provides a surprising fundamental improvement over all existent algorithms. It consists in applying, in parallel, the power iteration method with $m_\text{iter}$ iterations to $m_\text{init}$ different random initialization. Then, SMPI uses the maximum likelihood estimator to select the output vector in this subset by choosing the vector that maximizes $\tT ( \vv,\vv,\vv)$. It is described in the Algorithm 2.

\begin{algorithm}[h!]
  \caption{Selective Multiple Power Iteration}
    \label{algo:recoverysmpi}
\begin{algorithmic}[1]
  \State {\bfseries Input:} The tensor $\tT=\tZ + \beta \vv_0^{\otimes k}$, $m_\text{init}>10n$, $m_\text{iter}>10n$,$\Lambda$
  \State {\bfseries Goal:} Estimate $\vv_0$.
   
  \For{ i=0 to $m_\text{init}$} 
  \State Generate a random vector $\vv_{i,0}$
  \For{ j=0 to $m_\text{iter}$} 
  \State $\displaystyle\vv_{i,j+1}=\frac{\tT(:,\vv_{i,j},\vv_{i,j})}{\norm{\tT(:,\vv_{i,j},\vv_{i,j}) }}$
  \If{ $\displaystyle j>\Lambda$ and $ \abs{\langle \vv_{i,j-\Lambda},\vv_{i,j}\rangle}\geq 1-\varepsilon $}
  \State $\vv_{i,m_\text{iter}}=\vv_{i,j}$
  \State \textbf{break}
\EndIf
  \EndFor
\EndFor
  \State Select the vector $\vv=\arg \max_{1\leq i \leq {m_\text{init}}} \tT (\vv_{i,m_\text{iter}} ,\vv_{i,m_\text{iter}} ,\vv_{i,m_\text{iter}})$ 
  \State {\bfseries Output:} the estimated vector $\vv$
\end{algorithmic}
\end{algorithm}

 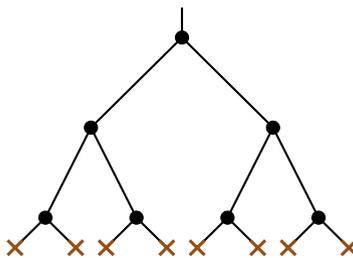
\begin{figure}[!h]
    \centering
\begin{tikzpicture}[>=latex,scale=0.8]
    \begin{scope}

    \coordinate (A) at (0,1.5) ;
    \draw[thick,-] (A) -- (-0.5,1)   ;
    \draw[thick,-] (A) -- (0.5,1)   ;
    
    \coordinate (A1) at (-1.5,1.5);
    \draw[thick,-] (A1) -- (-1,1)   ;
    \draw[thick,-] (A1)  -- (-2,1)   ;
    
    \coordinate (A2) at (-0.75,3);
    \draw[thick,-] (A2) -- (A)   ;
    \draw[thick,-] (A2) -- (A1)   ;
    
    \coordinate (AA) at (-3,1.5);
    \draw[thick,-] (AA) -- (-3.5,1)    ;
    \draw[thick,-] (AA) -- (-2.5,1)    ;
    
    \coordinate (AA1) at (-4.5,1.5) ;
    \draw[thick,-] (AA1) -- (-5,1)     ;
    \draw[thick,-] (AA1) -- (-4,1)     ;
    
    \coordinate (AA2) at (-3.75,3);
    \draw[thick,-] (AA2) -- (AA)   ;
    \draw[thick,-] (AA2) -- (AA1)   ;
    
    \coordinate (AAA) at (-2.25,4.5);
    \draw[thick,-] (AAA) -- (A2)  ;
    \draw[thick,-] (AAA) -- (AA2)  ;
    
    \draw[thick,-] (AAA) -- (-2.25,5)   ;

    \node[draw,circle,inner sep=1.75pt,fill] at (A){};
    \node[draw,circle,inner sep=1.75pt,fill] at (A1){};
    \node[draw,circle,inner sep=1.75pt,fill] at (A2){};
    \node[draw,circle,inner sep=1.75pt,fill] at (AA){};
    \node[draw,circle,inner sep=1.75pt,fill] at (AA1){};
    \node[draw,circle,inner sep=1.75pt,fill] at (AA2){};
    \node[draw,circle,inner sep=1.75pt,fill] at (AAA){};

    \draw (-0.5,1) node[cross=4pt,purple2] {};
    \draw (0.5,1) node[cross=4pt,purple2] {};
    \draw (-1,1) node[cross=4pt,purple2] {};
    \draw (-2,1) node[cross=4pt,purple2] {};
    \draw (-2.5,1) node[cross=4pt,purple2] {};
    \draw (-3.5,1) node[cross=4pt,purple2] {};
    \draw (-4,1) node[cross=4pt,purple2] {};
    \draw (-5,1) node[cross=4pt,purple2] {};

    \end{scope}
\end{tikzpicture}
    \caption{The graph associated to the power iteration method with 3 iterations for an initialization $\vv$. The cross represents the vector $\vv$ and the black dot the tensor $\tT$}
    \label{fig:Variance Matrix}
\end{figure}

Interestingly, Power Iteration could also be studied through tensor invariants, as it was already partially investigated in \cite{Evnin:2020ddw}. The associated graph is illustrated in Figure 6.

\subsubsection{Comparison with the state of the art.}
We plot in Figure 7 the correlation between the output of each algorithm and the signal vector; for SMPI (blue), TensorLy (TenLy, red) \cite{TensorLy:v20:18-277} which a recent practical Python package widely used and the State-of-the-art represented here by the Unfolding (Unf, green) \cite{richard2014statistical} and Homotopy-based (Hom, orange) \cite{homotopy17a} methods for four values of the dimension of each axe of the tensor ($n= 100, 200, 400$).

We observe that SMPI already outperforms existent algorithms for $n=100$ and the only one whose performance does not decrease as $n$ grows. This is remarkable as it is conjectured that the algorithmic threshold (the $\beta$ above which the algorithm is successful) scales as $n^{1/4}$ even if the information theory theory is at O(1) (this fact is denoted the statistical-computational gap).
\begin{figure*}[h!]
\centering
\begin{subfigure}{.5\textwidth}
    \centering
    \includegraphics[width=0.95\textwidth]{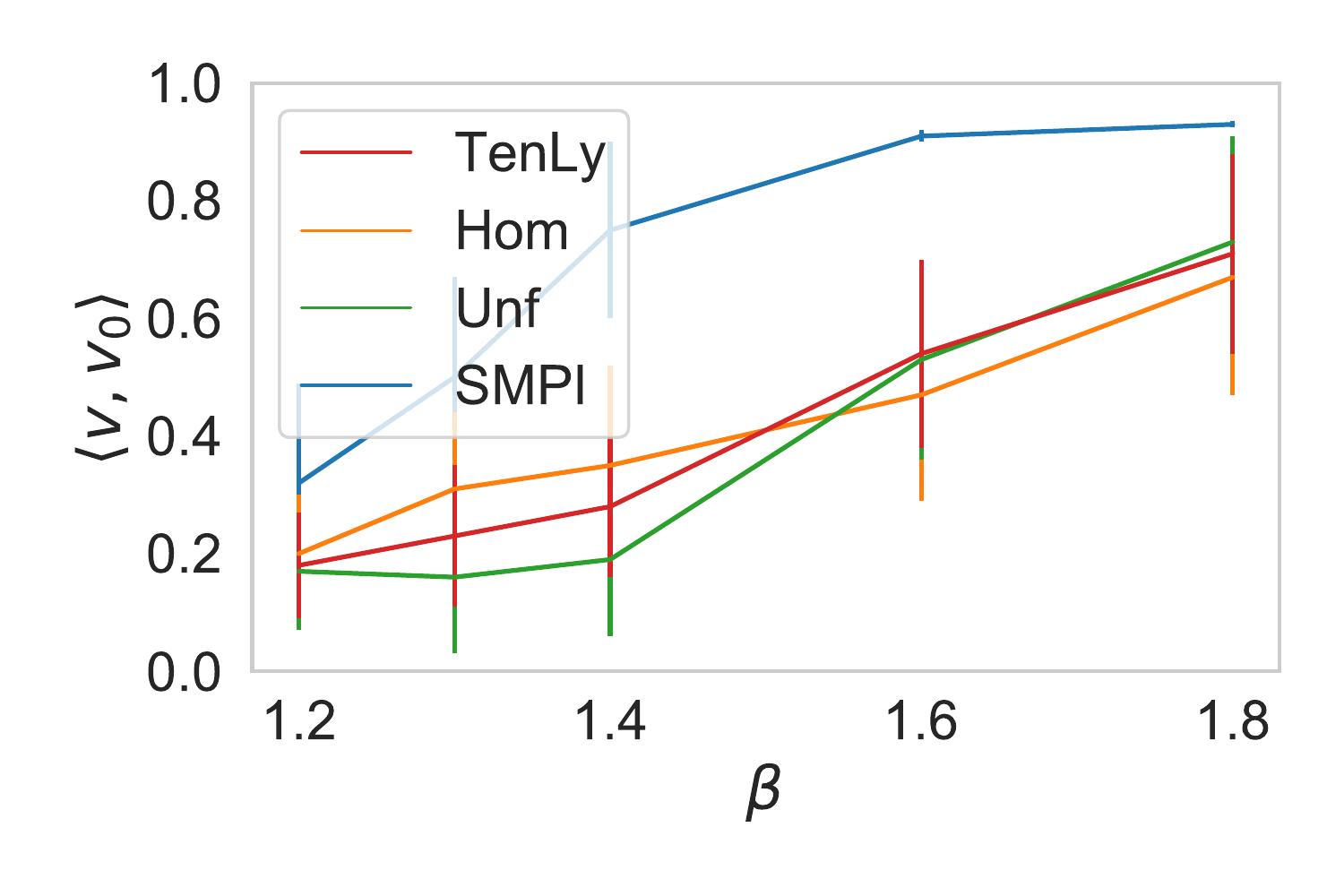}
    \label{fig:max93}
\end{subfigure}%
\begin{subfigure}{.5\textwidth}
    \centering
    \includegraphics[width=0.95\textwidth]{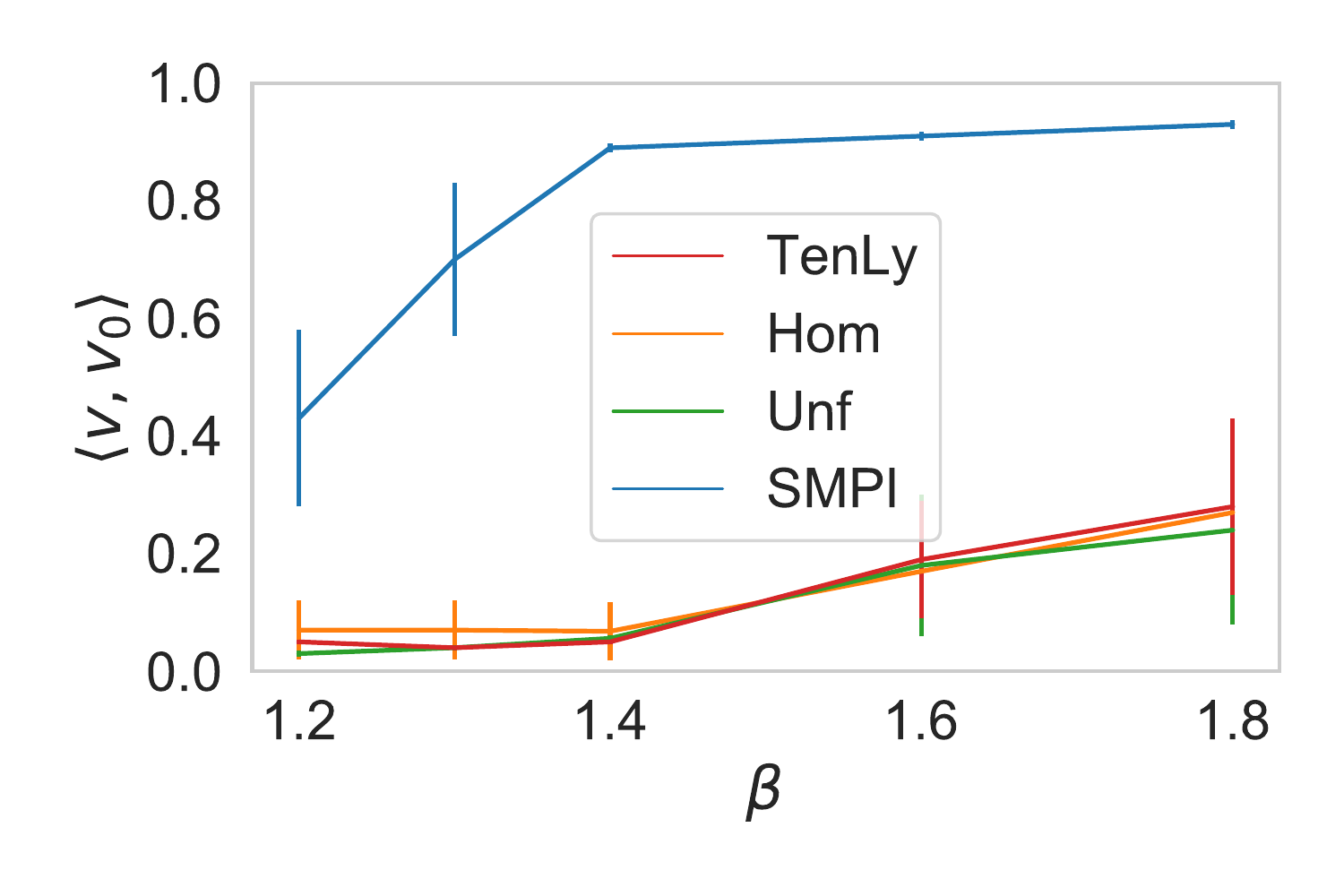}
    \label{fig:max2}
\end{subfigure}

\begin{subfigure}{.5\textwidth}
    \centering
    \includegraphics[width=0.95\textwidth]{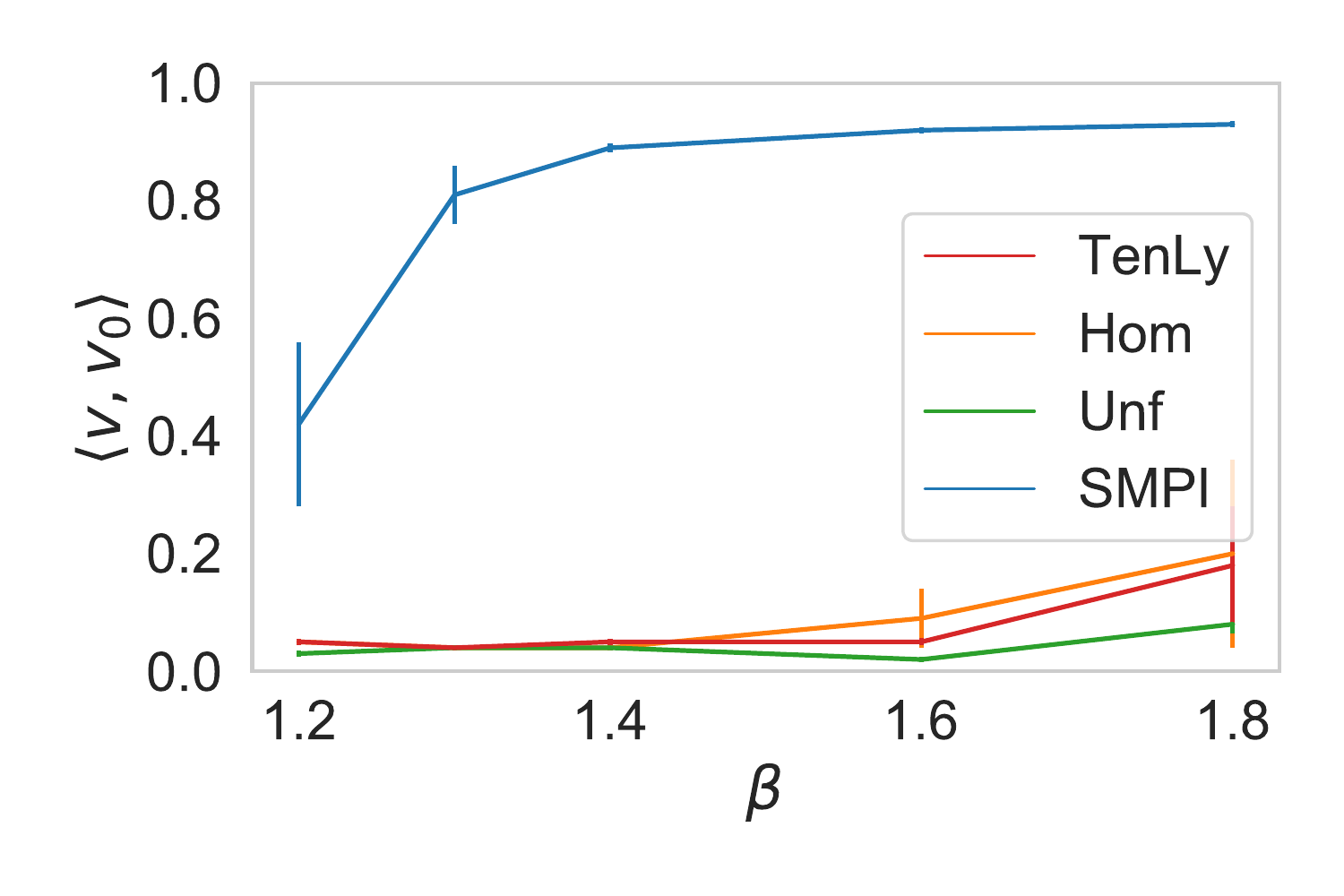}
    \label{fig:Per2}
\end{subfigure}

\caption{Comparison of the results of SMPI with TensorLy (TenLy) and the State-of-the-art represented here by the Unfolding (Unf) and Homotopy-based (Hom) methods for four values of the dimension of each axe of the tensor ($n= 100, 200, 400$). The results consist of the correlation between the output of each algorithm and the signal vector. }
\label{Comparison SOTA}
\end{figure*}

\subsubsection{Difference between SMPI and other power-iteration based algorithms}

Power iteration has been studied in various works with different settings. However, \cite{ouerfelli2021selective} experiments suggest that the success of SMPI is related to five essential non-trivial features. It turns out that, to the best of our knowledge, no other work studied Power Iteration with the five features simultaneously present, as summarized in the following Table 1. Thus, further theoretical investigations are required to unveil the theoretical insights behind this experimental success, which could lead to new important insights on the statistical-computational gap.

\begin{table}[h!]
\small
\setlength{\tabcolsep}{2.3pt}
\caption{The five essential features of SMPI compared to previous works investigating Power Iteration}
\label{tab:plateau}
\begin{center} \begin{tabular}{c|c|c|c|c|c}
\hline
  Algorithm &  Symmetry & \begin{tabular}{c}
       Discreet  \\
        step size
  \end{tabular}  & \begin{tabular}{c}
       Poly. nb  \\
        of initialisat.
  \end{tabular} &  \begin{tabular}{c}
       Poly. nb  \\
        of iterations
  \end{tabular} &\begin{tabular}{c}
      No stop  \\
        condition
  \end{tabular} \\
  \hline
  Wang et al., 2017 \cite{Wang_2017} &  Yes &  Yes & No &  No & No \\
  \hline
  Huang et al., 2020 \cite{huang2020power} &  No &  Yes & Yes &  Yes & Yes \\
  \hline
  Ben Arous et al., 2020 \cite{arous2020algorithmic} &  Yes &  No & Yes &  Yes & Yes \\
  \hline
  Dudeja et al., 2022 \cite{dudeja2022statistical} &  Yes &  Yes & Yes &  No & Yes \\
  \hline
  SMPI, 2021 \cite{ouerfelli2021selective} &  Yes &  Yes & Yes &  Yes & Yes \\
  \hline
   
\end{tabular}
\end{center}
\label{Tab:comp}
\end{table}

\section{Artificial Intelligence}


The importance of random matrices in Artificial Intelligence (AI) is now well-established. In view of this, random tensors and more generally, tensor tools are expected to play a growing role in many AI subjects. In this section, we give some examples of AI applications where tensors seem promising.  

\subsection{Application to image processing}
Image processing is an important topic in computer science due to its uncountable  applications. Since 2012, the performances in this domain have drastically increased with the revolution of deep learning \cite{krizhevsky2012imagenet}.
It is worth to note that deep learning have first demonstrated is strength on image processing (recognition of handwritten digits via the MNIST database \cite{lecun1998mnist}, large-scale image classification in the ImageNet database \cite{deng2009imagenet}, image synthesis using GAN architectures \cite{goodfellow2014generative},...).
However, it is also in this context that appeared one of the main weaknesses of deep learning, namely its sensitivity to small changes at the input level \cite{szegedy2013intriguing, nguyen2015deep, eykholt2018robust}. This problem is so important that it opens a completely new area of research known as adversarial examples \cite{yuan2019adversarial}. Indeed, these intriguing properties are clearly not suited for critical applications (involving human lives). This limitation of deep learning is amplified by its need of large amount of data, its enormous energy cost \cite{strubell2019energy, thompson2020computational} and its black box behavior \cite{castelvecchi2016can}, which all together limit the domain of applications for the deep learning approach.

\medskip
As a scientific challenge, it is important to find and develop alternative or complementary methods to deep learning models that are sufficiently powerful in order to deal robustly with image processing tasks, which is for example needed to address critical applications (e.g. autonomous vehicle). With this in mind, we argue that it would be interesting to develop and spread tensor methods for image processing in order to gain more robustness and theoretical guarantee in practical applications. Of course, the subject of tensor methods is not new in image processing \cite{panagakis2021tensor}. Indeed, its usefulness has already been shown for example for facial image representation \cite{vasilescu2002multilinear, tian2012multiview, tang2013tensor, lehky2020face}, for human motion analysis \cite{vasilescu2002human}, for classification of images of Handwritten digit \cite{savas2007handwritten}, for the general task of image denoising \cite{rajwade2012image} and for background-foreground separation in videos  \cite{anandkumar2016tensor, kajo2018svd, kajo2020tensor}.

\subsection{Application to Deep Neural Networks speed-up and compression}
Higher-order tensor clustering and tensor decompositions such as Tucker, Canonical Polyadic (CP) and Tensor Train (TT) decompositions have been successfully used for Convolutional Neural Networks (CNN) speed-up and compression in \cite{kim2015compression, wang2016accelerating} and also applied to fully connected neural networks \cite{novikov2015tensorizing} with an impressive achieved compression rate. However, even if these approaches have been addressed individually in multiple papers, as all tensor decompositions have been tried in principle, the combination of tensor decomposition with other approaches such as quantization \cite{hubara2017quantized} and distillation \cite{hinton2015distilling, romero2014fitnets, polino2018model, seddik2020lightweight} has not been well studied yet. Related to this combination-based approach, questions such as the following deserved to be explored: Is a “tensorized” neural network more suitable for the quantization than the original model?

\medskip
Another important research axis aims to take advantage of the recent algorithmic advances in the field of tensor methods \cite{ouerfelli2021selective, andriantsiory2021multi, Ouerfelli2022random}. Given the significant improvement of performance they achieved, it pushes us to rethink these new tools for the compression and speed up of Deep Neural Networks (DNN). It will be interesting to study the performance of these decompositions for a pre-trained neural network as well as during the training. For example, in the pre-trained case, there is already interesting aspect that we will investigate with respect to the fine-tuning procedure. Indeed, this approach usually starts with a pretrained CNN and then interleaves the decomposition of convolutional layers with fine-tuning operations \cite{lebedev2014speeding, astrid2017cp}. Each decomposition step that leads to a speed-up and compression, unfortunately comes with a drop in accuracy. Thus, the subsequent fine-tuning operation is expected to recover the accuracy drop. One important question is then, how these new tensor algorithms will reduce this drop in accuracy and will it still be necessary to apply the fine-tuning operation? 

\medskip
A third important research axis to investigate is related to the exploitation of the inherent sparsity present in DNN models \cite{liu2015sparse, wen2016learning}. To achieve this challenging goal, we need to develop Sparse Tensor Decomposition building on the new tensor methods findings utilizing insights from existent approaches  for sparse tensors \cite{allen2012sparse}. Closely related to this point, investigating and developing the Sparse tensor clustering will also be of great interest for the speed-up and compression of DNN.

\subsection{Application to particle trajectories}

In high energy physics, the matrix  principal component analysis  has been applied to the detection 
and recovery of particle trajectories emanating from accelerators. 
A major goal is to fit this MPCA to
the continuous flow of data from those accelerators, such as the many accelerators of CERN.

\medskip
On the other hand Tensor PCA is one essential tool for separating signal from noise.
The random tensor theories are making significant and continuous progress in the problems 
of data analysis. We reason that 
we have a good occasion to transform the subject and the corresponding algorithms.

\subsection{Other applications}
\medskip
More generally, it is important to note that there is many other practical applications where tensor decomposition and where SMPI could be an excellent candidate for improving existent performance. Here is a non-exhaustive list of such applications:

\medskip
Related to image processing, interesting applications of tensor methods have been explored in computer graphics. For example, a framework for image-based rendering dedicated to the realistic mapping of a texture onto a planar surface is presented in \cite{vasilescu2004tensortextures}. Another tensor-based application is the face transfer where a mapping is performed to apply a video recorded performances of one individual to facial animations of another \cite{vlasic2006face}   

\medskip
Hyperspectral images are naturally three-dimensional arrays so there have been different types of work exploiting tensor methods for such data. Tensor decomposition have been explored for different tasks such as hyperspectral image denoising \cite{liu2012denoising}, hyperspectral image super-resolution \cite{prevost2020hyperspectral}, hyperspectral image classification \cite{bourennane2010improvement}, for object detection \cite{bourennane2010improvementB} and also for medical hyperspectral image analysis \cite{dey2019tensor}.  

\medskip
In telecommunication, CP decomposition is used for tensor-based modulation \cite{decurninge2020tensor, decurninge2021tensor}. It used for massive random access, whereby a large number of transmitters communicate with a single receiver, ant it constitutes a key design challenge for future generations of wireless systems.


\medskip
Let us come to opportunities in the field of chemical and biology. For instance, in chemometrics \cite{henrion1994n} three-dimensional arrays may be generated by collecting data tables with a fixed set of objects and variables under different experimental conditions, at different sampling times, etc. One may also refer to the recent review \cite{sun2020opportunities} where different types of higher order data in manufacturing processes are presented. Their potential usage is addressed using methods like CP tensor decomposition. The authors also give concrete perspectives on the application of tensorial data analytics to these kind of processes.

\medskip
We finish this section by briefly mentioning applications where tensorial data analytics has been explored. This is for example the case in geophysics in the context of three-dimensional irregular seismic data reconstruction \cite{ma2013three}. Electroencephalogram (EEG) data collected during a cognitive control task may also benefit from tensor methods \cite{mahyari2016tensor}. Other interesting applications of tensor methods are dedicated to natural language processing \cite{sobhani2019text} and radar data \cite{nion2010tensor}. We can also mention the following works \cite{tichai2019tensor} concerning the application of tensor decomposition techniques to many-body tensors which has proven highly beneficial to reduce the computational cost in quantum chemistry and solid-state physics.

\section{Conclusion}

This review is a modest step in the direction of bring closer the different people working on random tensors
and it suggests research in many directions, among which:
\begin{itemize}
\item investigate better some tensorial fields models such as $T^4_D$ family 
and determine their renormalisation group flow,
\item 
understand better the connection between tensors-inspired algorithms and disordered systems, 

\item figure out the concrete  applications relevant to data analysis and to artificial intelligence.

\end{itemize}

\bibliography{sn-bibliography}


\begin{thebibliography}{103}
\ifx \bisbn   \undefined \def \bisbn  #1{ISBN #1}\fi
\ifx \binits  \undefined \def \binits#1{#1}\fi
\ifx \bauthor  \undefined \def \bauthor#1{#1}\fi
\ifx \batitle  \undefined \def \batitle#1{#1}\fi
\ifx \bjtitle  \undefined \def \bjtitle#1{#1}\fi
\ifx \bvolume  \undefined \def \bvolume#1{\textbf{#1}}\fi
\ifx \byear  \undefined \def \byear#1{#1}\fi
\ifx \bissue  \undefined \def \bissue#1{#1}\fi
\ifx \bfpage  \undefined \def \bfpage#1{#1}\fi
\ifx \blpage  \undefined \def \blpage #1{#1}\fi
\ifx \burl  \undefined \def \burl#1{\textsf{#1}}\fi
\ifx \doiurl  \undefined \def \doiurl#1{\url{https://doi.org/#1}}\fi
\ifx \betal  \undefined \def \betal{\textit{et al.}}\fi
\ifx \binstitute  \undefined \def \binstitute#1{#1}\fi
\ifx \binstitutionaled  \undefined \def \binstitutionaled#1{#1}\fi
\ifx \bctitle  \undefined \def \bctitle#1{#1}\fi
\ifx \beditor  \undefined \def \beditor#1{#1}\fi
\ifx \bpublisher  \undefined \def \bpublisher#1{#1}\fi
\ifx \bbtitle  \undefined \def \bbtitle#1{#1}\fi
\ifx \bedition  \undefined \def \bedition#1{#1}\fi
\ifx \bseriesno  \undefined \def \bseriesno#1{#1}\fi
\ifx \blocation  \undefined \def \blocation#1{#1}\fi
\ifx \bsertitle  \undefined \def \bsertitle#1{#1}\fi
\ifx \bsnm \undefined \def \bsnm#1{#1}\fi
\ifx \bsuffix \undefined \def \bsuffix#1{#1}\fi
\ifx \bparticle \undefined \def \bparticle#1{#1}\fi
\ifx \barticle \undefined \def \barticle#1{#1}\fi
\bibcommenthead
\ifx \bconfdate \undefined \def \bconfdate #1{#1}\fi
\ifx \botherref \undefined \def \botherref #1{#1}\fi
\ifx \url \undefined \def \url#1{\textsf{#1}}\fi
\ifx \bchapter \undefined \def \bchapter#1{#1}\fi
\ifx \bbook \undefined \def \bbook#1{#1}\fi
\ifx \bcomment \undefined \def \bcomment#1{#1}\fi
\ifx \oauthor \undefined \def \oauthor#1{#1}\fi
\ifx \citeauthoryear \undefined \def \citeauthoryear#1{#1}\fi
\ifx \endbibitem  \undefined \def \endbibitem {}\fi
\ifx \bconflocation  \undefined \def \bconflocation#1{#1}\fi
\ifx \arxivurl  \undefined \def \arxivurl#1{\textsf{#1}}\fi
\csname PreBibitemsHook\endcsname

\bibitem{t1993planar}
\begin{barticle}
\bauthor{\bsnm{Hooft}, \binits{G.t.}}:
\batitle{A planar diagram theory for strong interactions}.
\bjtitle{Nuclear Physics B}
\bvolume{72}(\bissue{3}),
\bfpage{461}--\blpage{473}
(\byear{1974})
\end{barticle}
\endbibitem

\bibitem{ambjorn1991three}
\begin{barticle}
\bauthor{\bsnm{Ambj{\o}rn}, \binits{J.}},
\bauthor{\bsnm{Durhuus}, \binits{B.}},
\bauthor{\bsnm{Jonsson}, \binits{T.}}:
\batitle{Three-dimensional simplicial quantum gravity and generalized matrix
  models}.
\bjtitle{Modern Physics Letters A}
\bvolume{6}(\bissue{12}),
\bfpage{1133}--\blpage{1146}
(\byear{1991})
\end{barticle}
\endbibitem

\bibitem{gross1992tensor}
\begin{barticle}
\bauthor{\bsnm{Gross}, \binits{M.}}:
\batitle{Tensor models and simplicial quantum gravity in> 2-d}.
\bjtitle{Nuclear Physics B-Proceedings Supplements}
\bvolume{25},
\bfpage{144}--\blpage{149}
(\byear{1992})
\end{barticle}
\endbibitem

\bibitem{sasakura1991tensor}
\begin{barticle}
\bauthor{\bsnm{Sasakura}, \binits{N.}}:
\batitle{Tensor model for gravity and orientability of manifold}.
\bjtitle{Modern Physics Letters A}
\bvolume{6}(\bissue{28}),
\bfpage{2613}--\blpage{2623}
(\byear{1991})
\end{barticle}
\endbibitem

\bibitem{gurau20111}
\begin{bchapter}
\bauthor{\bsnm{Gurau}, \binits{R.}}:
\bctitle{The $1/n$ expansion of colored tensor models}.
In: \bbtitle{Annales Henri Poincare},
vol. \bseriesno{12},
pp. \bfpage{829}--\blpage{847}
(\byear{2011}).
\bcomment{Springer}
\end{bchapter}
\endbibitem

\bibitem{guruau2017random}
\begin{bbook}
\bauthor{\bsnm{Gurau}, \binits{R.}}:
\bbtitle{Random Tensors}.
\bpublisher{Oxford University Press},
\blocation{Oxford}
(\byear{2017})
\end{bbook}
\endbibitem

\bibitem{gurau20111a}
\begin{barticle}
\bauthor{\bsnm{Gurau}, \binits{R.}},
\bauthor{\bsnm{Rivasseau}, \binits{V.}}:
\batitle{The $1/n$ expansion of colored tensor models in arbitrary dimension}.
\bjtitle{EPL (Europhysics Letters)}
\bvolume{95}(\bissue{5}),
\bfpage{50004}
(\byear{2011})
\end{barticle}
\endbibitem

\bibitem{rivasseau2012quantum}
\begin{bchapter}
\bauthor{\bsnm{Rivasseau}, \binits{V.}}:
\bctitle{Quantum gravity and renormalization: the tensor track}.
In: \bbtitle{AIP Conference Proceedings 8},
vol. \bseriesno{1444},
pp. \bfpage{18}--\blpage{29}
(\byear{2012}).
\bcomment{American Institute of Physics}
\end{bchapter}
\endbibitem

\bibitem{loll2019quantum}
\begin{barticle}
\bauthor{\bsnm{Loll}, \binits{R.}}:
\batitle{Quantum gravity from causal dynamical triangulations: a review}.
\bjtitle{Classical and Quantum Gravity}
\bvolume{37}(\bissue{1}),
\bfpage{013002}
(\byear{2019})
\end{barticle}
\endbibitem

\bibitem{szabo2003quantum}
\begin{barticle}
\bauthor{\bsnm{Szabo}, \binits{R.J.}}:
\batitle{Quantum field theory on noncommutative spaces}.
\bjtitle{Physics Reports}
\bvolume{378}(\bissue{4}),
\bfpage{207}--\blpage{299}
(\byear{2003})
\end{barticle}
\endbibitem

\bibitem{grosse2014self}
\begin{barticle}
\bauthor{\bsnm{Grosse}, \binits{H.}},
\bauthor{\bsnm{Wulkenhaar}, \binits{R.}}:
\batitle{Self-dual noncommutative $\phi^4$ theory in four dimensions is a
  non-perturbatively solvable and non-trivial quantum field theory}.
\bjtitle{Communications in Mathematical Physics}
\bvolume{329}(\bissue{3}),
\bfpage{1069}--\blpage{1130}
(\byear{2014})
\end{barticle}
\endbibitem

\bibitem{branahl2021scalar}
\begin{botherref}
\oauthor{\bsnm{Branahl}, \binits{J.}},
\oauthor{\bsnm{Grosse}, \binits{H.}},
\oauthor{\bsnm{Hock}, \binits{A.}},
\oauthor{\bsnm{Wulkenhaar}, \binits{R.}}:
From scalar fields on quantum spaces to blobbed topological recursion.
arXiv preprint arXiv:2110.11789
(2021)
\end{botherref}
\endbibitem

\bibitem{oriti2016group}
\begin{barticle}
\bauthor{\bsnm{Oriti}, \binits{D.}}:
\batitle{Group field theory as the second quantization of loop quantum
  gravity}.
\bjtitle{Classical and Quantum Gravity}
\bvolume{33}(\bissue{8}),
\bfpage{085005}
(\byear{2016})
\end{barticle}
\endbibitem

\bibitem{geloun2013renormalizable}
\begin{barticle}
\bauthor{\bsnm{Geloun}, \binits{J.B.}},
\bauthor{\bsnm{Rivasseau}, \binits{V.}}:
\batitle{A renormalizable 4-dimensional tensor field theory}.
\bjtitle{Communications in Mathematical Physics}
\bvolume{318}(\bissue{1}),
\bfpage{69}--\blpage{109}
(\byear{2013})
\end{barticle}
\endbibitem

\bibitem{ben20133d}
\begin{bchapter}
\bauthor{\bsnm{Ben~Geloun}, \binits{J.}},
\bauthor{\bsnm{Samary}, \binits{D.O.}}:
\bctitle{3d tensor field theory: renormalization and one-loop
  $\beta$-functions}.
In: \bbtitle{Annales Henri Poincare},
vol. \bseriesno{14},
pp. \bfpage{1599}--\blpage{1642}
(\byear{2013}).
\bcomment{Springer}
\end{bchapter}
\endbibitem

\bibitem{rivasseau2015tensor}
\begin{barticle}
\bauthor{\bsnm{Rivasseau}, \binits{V.}}:
\batitle{Why are tensor field theories asymptotically free?}
\bjtitle{EPL (Europhysics Letters)}
\bvolume{111}(\bissue{6}),
\bfpage{60011}
(\byear{2015})
\end{barticle}
\endbibitem

\bibitem{rivasseau2014tensor}
\begin{barticle}
\bauthor{\bsnm{Rivasseau}, \binits{V.}}:
\batitle{The tensor track, iii}.
\bjtitle{Fortschritte der Physik}
\bvolume{62}(\bissue{2}),
\bfpage{81}--\blpage{107}
(\byear{2014})
\end{barticle}
\endbibitem

\bibitem{delporte2018tensor}
\begin{botherref}
\oauthor{\bsnm{Delporte}, \binits{N.}},
\oauthor{\bsnm{Rivasseau}, \binits{V.}}:
The tensor track v: holographic tensors.
arXiv preprint arXiv:1804.11101
(2018)
\end{botherref}
\endbibitem

\bibitem{delporte2020tensor}
\begin{botherref}
\oauthor{\bsnm{Delporte}, \binits{N.}},
\oauthor{\bsnm{Rivasseau}, \binits{V.}}:
The tensor track vi: Field theory on random trees and syk on random unicyclic
  graphs.
arXiv preprint arXiv:2004.13744
(2020)
\end{botherref}
\endbibitem

\bibitem{bonzom2012random}
\begin{barticle}
\bauthor{\bsnm{Bonzom}, \binits{V.}},
\bauthor{\bsnm{Gurau}, \binits{R.}},
\bauthor{\bsnm{Rivasseau}, \binits{V.}}:
\batitle{Random tensor models in the large n limit: Uncoloring the colored
  tensor models}.
\bjtitle{Physical Review D}
\bvolume{85}(\bissue{8}),
\bfpage{084037}
(\byear{2012})
\end{barticle}
\endbibitem

\bibitem{bonzom2011critical}
\begin{barticle}
\bauthor{\bsnm{Bonzom}, \binits{V.}},
\bauthor{\bsnm{Gurau}, \binits{R.}},
\bauthor{\bsnm{Riello}, \binits{A.}},
\bauthor{\bsnm{Rivasseau}, \binits{V.}}:
\batitle{Critical behavior of colored tensor models in the large n limit}.
\bjtitle{Nuclear Physics B}
\bvolume{853}(\bissue{1}),
\bfpage{174}--\blpage{195}
(\byear{2011})
\end{barticle}
\endbibitem

\bibitem{gurau2014melons}
\begin{bchapter}
\bauthor{\bsnm{Gurau}, \binits{R.}},
\bauthor{\bsnm{Ryan}, \binits{J.P.}}:
\bctitle{Melons are branched polymers}.
In: \bbtitle{Annales Henri Poincar{\'e}},
vol. \bseriesno{15},
pp. \bfpage{2085}--\blpage{2131}
(\byear{2014}).
\bcomment{Springer}
\end{bchapter}
\endbibitem

\bibitem{gurau2016regular}
\begin{barticle}
\bauthor{\bsnm{Gurau}, \binits{R.G.}},
\bauthor{\bsnm{Schaeffer}, \binits{G.}}:
\batitle{Regular colored graphs of positive degree}.
\bjtitle{Annales de l’Institut Henri Poincar{\'e} D}
\bvolume{3}(\bissue{3}),
\bfpage{257}--\blpage{320}
(\byear{2016})
\end{barticle}
\endbibitem

\bibitem{dartois2013double}
\begin{barticle}
\bauthor{\bsnm{Dartois}, \binits{S.}},
\bauthor{\bsnm{Gurau}, \binits{R.}},
\bauthor{\bsnm{Rivasseau}, \binits{V.}}:
\batitle{Double scaling in tensor models with a quartic interaction}.
\bjtitle{Journal of High Energy Physics}
\bvolume{2013}(\bissue{9}),
\bfpage{1}--\blpage{33}
(\byear{2013})
\end{barticle}
\endbibitem

\bibitem{lionni2021iterated}
\begin{barticle}
\bauthor{\bsnm{Lionni}, \binits{L.}},
\bauthor{\bsnm{Marckert}, \binits{J.-F.}}:
\batitle{Iterated foldings of discrete spaces and their limits: Candidates for
  the role of brownian map in higher cimensions}.
\bjtitle{Mathematical Physics, Analysis and Geometry}
\bvolume{24}(\bissue{4}),
\bfpage{1}--\blpage{67}
(\byear{2021})
\end{barticle}
\endbibitem

\bibitem{maldacena2016remarks}
\begin{barticle}
\bauthor{\bsnm{Maldacena}, \binits{J.}},
\bauthor{\bsnm{Stanford}, \binits{D.}}:
\batitle{Remarks on the sachdev-ye-kitaev model}.
\bjtitle{Physical Review D}
\bvolume{94}(\bissue{10}),
\bfpage{106002}
(\byear{2016})
\end{barticle}
\endbibitem

\bibitem{gross2017generalization}
\begin{barticle}
\bauthor{\bsnm{Gross}, \binits{D.J.}},
\bauthor{\bsnm{Rosenhaus}, \binits{V.}}:
\batitle{A generalization of sachdev-ye-kitaev}.
\bjtitle{Journal of High Energy Physics}
\bvolume{2017}(\bissue{2}),
\bfpage{1}--\blpage{38}
(\byear{2017})
\end{barticle}
\endbibitem

\bibitem{maldacena2016bound}
\begin{barticle}
\bauthor{\bsnm{Maldacena}, \binits{J.}},
\bauthor{\bsnm{Shenker}, \binits{S.H.}},
\bauthor{\bsnm{Stanford}, \binits{D.}}:
\batitle{A bound on chaos}.
\bjtitle{Journal of High Energy Physics}
\bvolume{2016}(\bissue{8}),
\bfpage{1}--\blpage{17}
(\byear{2016})
\end{barticle}
\endbibitem

\bibitem{witten2019syk}
\begin{barticle}
\bauthor{\bsnm{Witten}, \binits{E.}}:
\batitle{An syk-like model without disorder}.
\bjtitle{Journal of Physics A: Mathematical and Theoretical}
\bvolume{52}(\bissue{47}),
\bfpage{474002}
(\byear{2019})
\end{barticle}
\endbibitem

\bibitem{klebanov2017uncolored}
\begin{barticle}
\bauthor{\bsnm{Klebanov}, \binits{I.R.}},
\bauthor{\bsnm{Tarnopolsky}, \binits{G.}}:
\batitle{Uncolored random tensors, melon diagrams, and the sachdev-ye-kitaev
  models}.
\bjtitle{Physical Review D}
\bvolume{95}(\bissue{4}),
\bfpage{046004}
(\byear{2017})
\end{barticle}
\endbibitem

\bibitem{carrozza2016n}
\begin{barticle}
\bauthor{\bsnm{Carrozza}, \binits{S.}},
\bauthor{\bsnm{Tanasa}, \binits{A.}}:
\batitle{O (n) random tensor models}.
\bjtitle{Letters in Mathematical Physics}
\bvolume{106}(\bissue{11}),
\bfpage{1531}--\blpage{1559}
(\byear{2016})
\end{barticle}
\endbibitem

\bibitem{benedetti20191}
\begin{barticle}
\bauthor{\bsnm{Benedetti}, \binits{D.}},
\bauthor{\bsnm{Carrozza}, \binits{S.}},
\bauthor{\bsnm{Gurau}, \binits{R.}},
\bauthor{\bsnm{Kolanowski}, \binits{M.}}:
\batitle{The 1/n expansion of the symmetric traceless and the antisymmetric
  tensor models in rank three}.
\bjtitle{Communications in Mathematical Physics}
\bvolume{371}(\bissue{1}),
\bfpage{55}--\blpage{97}
(\byear{2019})
\end{barticle}
\endbibitem

\bibitem{carrozza2019syk}
\begin{barticle}
\bauthor{\bsnm{Carrozza}, \binits{S.}},
\bauthor{\bsnm{Pozsgay}, \binits{V.}}:
\batitle{Syk-like tensor quantum mechanics with sp (n) symmetry}.
\bjtitle{Nuclear Physics B}
\bvolume{941},
\bfpage{28}--\blpage{52}
(\byear{2019})
\end{barticle}
\endbibitem

\bibitem{carrozza2022melonic}
\begin{botherref}
\oauthor{\bsnm{Carrozza}, \binits{S.}},
\oauthor{\bsnm{Harribey}, \binits{S.}}:
Melonic large n limit of 5-index irreducible random tensors.
Communications in Mathematical Physics,
1--52
(2022)
\end{botherref}
\endbibitem

\bibitem{richard2014statistical}
\begin{bchapter}
\bauthor{\bsnm{Richard}, \binits{E.}},
\bauthor{\bsnm{Montanari}, \binits{A.}}:
\bctitle{A statistical model for tensor pca}.
In: \bbtitle{Advances in Neural Information Processing Systems},
pp. \bfpage{2897}--\blpage{2905}
(\byear{2014})
\end{bchapter}
\endbibitem

\bibitem{vasilescu2002multilinear}
\begin{bchapter}
\bauthor{\bsnm{Vasilescu}, \binits{M.A.O.}},
\bauthor{\bsnm{Terzopoulos}, \binits{D.}}:
\bctitle{Multilinear analysis of image ensembles: Tensorfaces}.
In: \bbtitle{European Conference on Computer Vision},
pp. \bfpage{447}--\blpage{460}
(\byear{2002}).
\bcomment{Springer}
\end{bchapter}
\endbibitem

\bibitem{nasrabadi2013hyperspectral}
\begin{barticle}
\bauthor{\bsnm{Nasrabadi}, \binits{N.M.}}:
\batitle{Hyperspectral target detection: An overview of current and future
  challenges}.
\bjtitle{IEEE Signal Processing Magazine}
\bvolume{31}(\bissue{1}),
\bfpage{34}--\blpage{44}
(\byear{2013})
\end{barticle}
\endbibitem

\bibitem{astrid2017cp}
\begin{bchapter}
\bauthor{\bsnm{Astrid}, \binits{M.}},
\bauthor{\bsnm{Lee}, \binits{S.-I.}}:
\bctitle{Cp-decomposition with tensor power method for convolutional neural
  networks compression}.
In: \bbtitle{2017 IEEE International Conference on Big Data and Smart Computing
  (BigComp)},
pp. \bfpage{115}--\blpage{118}
(\byear{2017}).
\bcomment{IEEE}
\end{bchapter}
\endbibitem

\bibitem{wang2020cpac}
\begin{botherref}
\oauthor{\bsnm{Wang}, \binits{Y.}},
\oauthor{\bsnm{Yue}, \binits{X.}}, et al.:
Cpac-conv: Cp-decomposition to approximately compress convolutional layers in
  deep learning.
arXiv preprint arXiv:2005.13746
(2020)
\end{botherref}
\endbibitem

\bibitem{lahat2015multimodal}
\begin{barticle}
\bauthor{\bsnm{Lahat}, \binits{D.}},
\bauthor{\bsnm{Adali}, \binits{T.}},
\bauthor{\bsnm{Jutten}, \binits{C.}}:
\batitle{Multimodal data fusion: an overview of methods, challenges, and
  prospects}.
\bjtitle{Proceedings of the IEEE}
\bvolume{103}(\bissue{9}),
\bfpage{1449}--\blpage{1477}
(\byear{2015})
\end{barticle}
\endbibitem

\bibitem{decurninge2020tensor}
\begin{botherref}
\oauthor{\bsnm{Decurninge}, \binits{A.}},
\oauthor{\bsnm{Land}, \binits{I.}},
\oauthor{\bsnm{Guillaud}, \binits{M.}}:
Tensor-based modulation for unsourced massive random access.
IEEE Wireless Communications Letters
(2020)
\end{botherref}
\endbibitem

\bibitem{panagakis2021tensor}
\begin{barticle}
\bauthor{\bsnm{Panagakis}, \binits{Y.}},
\bauthor{\bsnm{Kossaifi}, \binits{J.}},
\bauthor{\bsnm{Chrysos}, \binits{G.G.}},
\bauthor{\bsnm{Oldfield}, \binits{J.}},
\bauthor{\bsnm{Nicolaou}, \binits{M.A.}},
\bauthor{\bsnm{Anandkumar}, \binits{A.}},
\bauthor{\bsnm{Zafeiriou}, \binits{S.}}:
\batitle{Tensor methods in computer vision and deep learning}.
\bjtitle{Proceedings of the IEEE}
\bvolume{109}(\bissue{5}),
\bfpage{863}--\blpage{890}
(\byear{2021})
\end{barticle}
\endbibitem

\bibitem{sobhani2019text}
\begin{bchapter}
\bauthor{\bsnm{Sobhani}, \binits{E.}},
\bauthor{\bsnm{Comon}, \binits{P.}},
\bauthor{\bsnm{Jutten}, \binits{C.}},
\bauthor{\bsnm{Babaie-Zadeh}, \binits{M.}}:
\bctitle{Text mining with constrained tensor decomposition}.
In: \bbtitle{International Conference on Machine Learning, Optimization, and
  Data Science},
pp. \bfpage{219}--\blpage{231}
(\byear{2019}).
\bcomment{Springer}
\end{bchapter}
\endbibitem

\bibitem{arous2020algorithmic}
\begin{barticle}
\bauthor{\bsnm{Arous}, \binits{G.B.}},
\bauthor{\bsnm{Gheissari}, \binits{R.}},
\bauthor{\bsnm{Jagannath}, \binits{A.}}, \betal:
\batitle{Algorithmic thresholds for tensor pca}.
\bjtitle{Annals of Probability}
\bvolume{48}(\bissue{4}),
\bfpage{2052}--\blpage{2087}
(\byear{2020})
\end{barticle}
\endbibitem

\bibitem{mannelli2019passed}
\begin{bchapter}
\bauthor{\bsnm{Mannelli}, \binits{S.S.}},
\bauthor{\bsnm{Krzakala}, \binits{F.}},
\bauthor{\bsnm{Urbani}, \binits{P.}},
\bauthor{\bsnm{Zdeborova}, \binits{L.}}:
\bctitle{Passed \& spurious: Descent algorithms and local minima in spiked
  matrix-tensor models}.
In: \bbtitle{International Conference on Machine Learning},
pp. \bfpage{4333}--\blpage{4342}
(\byear{2019}).
\bcomment{PMLR}
\end{bchapter}
\endbibitem

\bibitem{mannelli2019afraid}
\begin{botherref}
\oauthor{\bsnm{Mannelli}, \binits{S.S.}},
\oauthor{\bsnm{Biroli}, \binits{G.}},
\oauthor{\bsnm{Cammarota}, \binits{C.}},
\oauthor{\bsnm{Krzakala}, \binits{F.}},
\oauthor{\bsnm{Zdeborov{\'a}}, \binits{L.}}:
Who is afraid of big bad minima? analysis of gradient-flow in a spiked
  matrix-tensor model.
arXiv preprint arXiv:1907.08226
(2019)
\end{botherref}
\endbibitem

\bibitem{mannelli2020marvels}
\begin{barticle}
\bauthor{\bsnm{Mannelli}, \binits{S.S.}},
\bauthor{\bsnm{Biroli}, \binits{G.}},
\bauthor{\bsnm{Cammarota}, \binits{C.}},
\bauthor{\bsnm{Krzakala}, \binits{F.}},
\bauthor{\bsnm{Urbani}, \binits{P.}},
\bauthor{\bsnm{Zdeborov{\'a}}, \binits{L.}}:
\batitle{Marvels and pitfalls of the langevin algorithm in noisy
  high-dimensional inference}.
\bjtitle{Physical Review X}
\bvolume{10}(\bissue{1}),
\bfpage{011057}
(\byear{2020})
\end{barticle}
\endbibitem

\bibitem{luo2020open}
\begin{bchapter}
\bauthor{\bsnm{Luo}, \binits{Y.}},
\bauthor{\bsnm{Zhang}, \binits{A.R.}}:
\bctitle{Open problem: Average-case hardness of hypergraphic planted clique
  detection}.
In: \bbtitle{Conference on Learning Theory},
pp. \bfpage{3852}--\blpage{3856}
(\byear{2020}).
\bcomment{PMLR}
\end{bchapter}
\endbibitem

\bibitem{Ouerfelli2022random}
\begin{bchapter}
\bauthor{\bsnm{Ouerfelli}, \binits{M.}},
\bauthor{\bsnm{Tamaazousti}, \binits{M.}},
\bauthor{\bsnm{Rivasseau}, \binits{V.}}:
\bctitle{Random tensor theory for tensor decomposition}.
In: \bbtitle{Proceedings of the AAAI Conference on Artificial Intelligence}
(\byear{2022})
\end{bchapter}
\endbibitem

\bibitem{ouerfelli2021selective}
\begin{botherref}
\oauthor{\bsnm{Ouerfelli}, \binits{M.}},
\oauthor{\bsnm{Tamaazousti}, \binits{M.}},
\oauthor{\bsnm{Rivasseau}, \binits{V.}}:
Selective Multiple Power Iteration: from Tensor PCA to gradient-based
  exploration of landscapes
(2021)
\end{botherref}
\endbibitem

\bibitem{Evnin:2020ddw}
\begin{botherref}
\oauthor{\bsnm{Evnin}, \binits{O.}}:
Melonic dominance and the largest eigenvalue of a large random tensor.
arXiv preprint arXiv:2003.11220
(2020)
\end{botherref}
\endbibitem

\bibitem{TensorLy:v20:18-277}
\begin{barticle}
\bauthor{\bsnm{Kossaifi}, \binits{J.}},
\bauthor{\bsnm{Panagakis}, \binits{Y.}},
\bauthor{\bsnm{Anandkumar}, \binits{A.}},
\bauthor{\bsnm{Pantic}, \binits{M.}}:
\batitle{Tensorly: Tensor learning in python}.
\bjtitle{Journal of Machine Learning Research}
\bvolume{20}(\bissue{26}),
\bfpage{1}--\blpage{6}
(\byear{2019})
\end{barticle}
\endbibitem

\bibitem{homotopy17a}
\begin{bchapter}
\bauthor{\bsnm{Anandkumar}, \binits{A.}},
\bauthor{\bsnm{Deng}, \binits{Y.}},
\bauthor{\bsnm{Ge}, \binits{R.}},
\bauthor{\bsnm{Mobahi}, \binits{H.}}:
\bctitle{Homotopy analysis for tensor pca}.
In: \beditor{\bsnm{Kale}, \binits{S.}},
\beditor{\bsnm{Shamir}, \binits{O.}} (eds.)
\bbtitle{Proceedings of the 2017 Conference on Learning Theory}.
\bsertitle{Proceedings of Machine Learning Research},
vol. \bseriesno{65},
pp. \bfpage{79}--\blpage{104}.
\bpublisher{PMLR},
\blocation{U.S.}
(\byear{2017})
\end{bchapter}
\endbibitem

\bibitem{Wang_2017}
\begin{barticle}
\bauthor{\bsnm{Wang}, \binits{M.}},
\bauthor{\bsnm{Dao~Duc}, \binits{K.}},
\bauthor{\bsnm{Fischer}, \binits{J.}},
\bauthor{\bsnm{Song}, \binits{Y.S.}}:
\batitle{Operator norm inequalities between tensor unfoldings on the partition
  lattice}.
\bjtitle{Linear Algebra and its Applications}
\bvolume{520},
\bfpage{44}--\blpage{66}
(\byear{2017})
\end{barticle}
\endbibitem

\bibitem{huang2020power}
\begin{botherref}
\oauthor{\bsnm{Huang}, \binits{J.}},
\oauthor{\bsnm{Huang}, \binits{D.Z.}},
\oauthor{\bsnm{Yang}, \binits{Q.}},
\oauthor{\bsnm{Cheng}, \binits{G.}}:
Power iteration for tensor pca.
arXiv preprint arXiv:2012.13669
(2020)
\end{botherref}
\endbibitem

\bibitem{dudeja2022statistical}
\begin{botherref}
\oauthor{\bsnm{Dudeja}, \binits{R.}},
\oauthor{\bsnm{Hsu}, \binits{D.}}:
Statistical-computational trade-offs in tensor pca and related problems via
  communication complexity.
arXiv preprint arXiv:2204.07526
(2022)
\end{botherref}
\endbibitem

\bibitem{krizhevsky2012imagenet}
\begin{barticle}
\bauthor{\bsnm{Krizhevsky}, \binits{A.}},
\bauthor{\bsnm{Sutskever}, \binits{I.}},
\bauthor{\bsnm{Hinton}, \binits{G.E.}}:
\batitle{Imagenet classification with deep convolutional neural networks}.
\bjtitle{Advances in neural information processing systems}
\bvolume{25},
\bfpage{1097}--\blpage{1105}
(\byear{2012})
\end{barticle}
\endbibitem

\bibitem{lecun1998mnist}
\begin{botherref}
\oauthor{\bsnm{LeCun}, \binits{Y.}}:
The mnist database of handwritten digits.
http://yann. lecun. com/exdb/mnist/
(1998)
\end{botherref}
\endbibitem

\bibitem{deng2009imagenet}
\begin{bchapter}
\bauthor{\bsnm{Deng}, \binits{J.}},
\bauthor{\bsnm{Dong}, \binits{W.}},
\bauthor{\bsnm{Socher}, \binits{R.}},
\bauthor{\bsnm{Li}, \binits{L.-J.}},
\bauthor{\bsnm{Li}, \binits{K.}},
\bauthor{\bsnm{Fei-Fei}, \binits{L.}}:
\bctitle{Imagenet: A large-scale hierarchical image database}.
In: \bbtitle{2009 IEEE Conference on Computer Vision and Pattern Recognition},
pp. \bfpage{248}--\blpage{255}
(\byear{2009}).
\bcomment{Ieee}
\end{bchapter}
\endbibitem

\bibitem{goodfellow2014generative}
\begin{botherref}
\oauthor{\bsnm{Goodfellow}, \binits{I.}},
\oauthor{\bsnm{Pouget-Abadie}, \binits{J.}},
\oauthor{\bsnm{Mirza}, \binits{M.}},
\oauthor{\bsnm{Xu}, \binits{B.}},
\oauthor{\bsnm{Warde-Farley}, \binits{D.}},
\oauthor{\bsnm{Ozair}, \binits{S.}},
\oauthor{\bsnm{Courville}, \binits{A.}},
\oauthor{\bsnm{Bengio}, \binits{Y.}}:
Generative adversarial nets.
Advances in neural information processing systems
\textbf{27}
(2014)
\end{botherref}
\endbibitem

\bibitem{szegedy2013intriguing}
\begin{botherref}
\oauthor{\bsnm{Szegedy}, \binits{C.}},
\oauthor{\bsnm{Zaremba}, \binits{W.}},
\oauthor{\bsnm{Sutskever}, \binits{I.}},
\oauthor{\bsnm{Bruna}, \binits{J.}},
\oauthor{\bsnm{Erhan}, \binits{D.}},
\oauthor{\bsnm{Goodfellow}, \binits{I.}},
\oauthor{\bsnm{Fergus}, \binits{R.}}:
Intriguing properties of neural networks.
arXiv preprint arXiv:1312.6199
(2013)
\end{botherref}
\endbibitem

\bibitem{nguyen2015deep}
\begin{bchapter}
\bauthor{\bsnm{Nguyen}, \binits{A.}},
\bauthor{\bsnm{Yosinski}, \binits{J.}},
\bauthor{\bsnm{Clune}, \binits{J.}}:
\bctitle{Deep neural networks are easily fooled: High confidence predictions
  for unrecognizable images}.
In: \bbtitle{Proceedings of the IEEE Conference on Computer Vision and Pattern
  Recognition},
pp. \bfpage{427}--\blpage{436}
(\byear{2015})
\end{bchapter}
\endbibitem

\bibitem{eykholt2018robust}
\begin{bchapter}
\bauthor{\bsnm{Eykholt}, \binits{K.}},
\bauthor{\bsnm{Evtimov}, \binits{I.}},
\bauthor{\bsnm{Fernandes}, \binits{E.}},
\bauthor{\bsnm{Li}, \binits{B.}},
\bauthor{\bsnm{Rahmati}, \binits{A.}},
\bauthor{\bsnm{Xiao}, \binits{C.}},
\bauthor{\bsnm{Prakash}, \binits{A.}},
\bauthor{\bsnm{Kohno}, \binits{T.}},
\bauthor{\bsnm{Song}, \binits{D.}}:
\bctitle{Robust physical-world attacks on deep learning visual classification}.
In: \bbtitle{Proceedings of the IEEE Conference on Computer Vision and Pattern
  Recognition},
pp. \bfpage{1625}--\blpage{1634}
(\byear{2018})
\end{bchapter}
\endbibitem

\bibitem{yuan2019adversarial}
\begin{barticle}
\bauthor{\bsnm{Yuan}, \binits{X.}},
\bauthor{\bsnm{He}, \binits{P.}},
\bauthor{\bsnm{Zhu}, \binits{Q.}},
\bauthor{\bsnm{Li}, \binits{X.}}:
\batitle{Adversarial examples: Attacks and defenses for deep learning}.
\bjtitle{IEEE transactions on neural networks and learning systems}
\bvolume{30}(\bissue{9}),
\bfpage{2805}--\blpage{2824}
(\byear{2019})
\end{barticle}
\endbibitem

\bibitem{strubell2019energy}
\begin{botherref}
\oauthor{\bsnm{Strubell}, \binits{E.}},
\oauthor{\bsnm{Ganesh}, \binits{A.}},
\oauthor{\bsnm{McCallum}, \binits{A.}}:
Energy and policy considerations for deep learning in nlp.
arXiv preprint arXiv:1906.02243
(2019)
\end{botherref}
\endbibitem

\bibitem{thompson2020computational}
\begin{botherref}
\oauthor{\bsnm{Thompson}, \binits{N.C.}},
\oauthor{\bsnm{Greenewald}, \binits{K.}},
\oauthor{\bsnm{Lee}, \binits{K.}},
\oauthor{\bsnm{Manso}, \binits{G.F.}}:
The computational limits of deep learning.
arXiv preprint arXiv:2007.05558
(2020)
\end{botherref}
\endbibitem

\bibitem{castelvecchi2016can}
\begin{barticle}
\bauthor{\bsnm{Castelvecchi}, \binits{D.}}:
\batitle{Can we open the black box of ai?}
\bjtitle{Nature News}
\bvolume{538}(\bissue{7623}),
\bfpage{20}
(\byear{2016})
\end{barticle}
\endbibitem

\bibitem{tian2012multiview}
\begin{barticle}
\bauthor{\bsnm{Tian}, \binits{C.}},
\bauthor{\bsnm{Fan}, \binits{G.}},
\bauthor{\bsnm{Gao}, \binits{X.}},
\bauthor{\bsnm{Tian}, \binits{Q.}}:
\batitle{Multiview face recognition: from tensorface to v-tensorface and
  k-tensorface}.
\bjtitle{IEEE Transactions on Systems, Man, and Cybernetics, Part B
  (Cybernetics)}
\bvolume{42}(\bissue{2}),
\bfpage{320}--\blpage{333}
(\byear{2012})
\end{barticle}
\endbibitem

\bibitem{tang2013tensor}
\begin{bchapter}
\bauthor{\bsnm{Tang}, \binits{Y.}},
\bauthor{\bsnm{Salakhutdinov}, \binits{R.}},
\bauthor{\bsnm{Hinton}, \binits{G.}}:
\bctitle{Tensor analyzers}.
In: \bbtitle{International Conference on Machine Learning},
pp. \bfpage{163}--\blpage{171}
(\byear{2013}).
\bcomment{PMLR}
\end{bchapter}
\endbibitem

\bibitem{lehky2020face}
\begin{barticle}
\bauthor{\bsnm{Lehky}, \binits{S.R.}},
\bauthor{\bsnm{Phan}, \binits{A.H.}},
\bauthor{\bsnm{Cichocki}, \binits{A.}},
\bauthor{\bsnm{Tanaka}, \binits{K.}}:
\batitle{Face representations via tensorfaces of various complexities}.
\bjtitle{Neural Computation}
\bvolume{32}(\bissue{2}),
\bfpage{281}--\blpage{329}
(\byear{2020})
\end{barticle}
\endbibitem

\bibitem{vasilescu2002human}
\begin{bchapter}
\bauthor{\bsnm{Vasilescu}, \binits{M.A.O.}}:
\bctitle{Human motion signatures: Analysis, synthesis, recognition}.
In: \bbtitle{2002 International Conference on Pattern Recognition},
vol. \bseriesno{3},
pp. \bfpage{456}--\blpage{460}
(\byear{2002}).
\bcomment{IEEE}
\end{bchapter}
\endbibitem

\bibitem{savas2007handwritten}
\begin{barticle}
\bauthor{\bsnm{Savas}, \binits{B.}},
\bauthor{\bsnm{Eld{\'e}n}, \binits{L.}}:
\batitle{Handwritten digit classification using higher order singular value
  decomposition}.
\bjtitle{Pattern recognition}
\bvolume{40}(\bissue{3}),
\bfpage{993}--\blpage{1003}
(\byear{2007})
\end{barticle}
\endbibitem

\bibitem{rajwade2012image}
\begin{barticle}
\bauthor{\bsnm{Rajwade}, \binits{A.}},
\bauthor{\bsnm{Rangarajan}, \binits{A.}},
\bauthor{\bsnm{Banerjee}, \binits{A.}}:
\batitle{Image denoising using the higher order singular value decomposition}.
\bjtitle{IEEE Transactions on Pattern Analysis and Machine Intelligence}
\bvolume{35}(\bissue{4}),
\bfpage{849}--\blpage{862}
(\byear{2012})
\end{barticle}
\endbibitem

\bibitem{anandkumar2016tensor}
\begin{bchapter}
\bauthor{\bsnm{Anandkumar}, \binits{A.}},
\bauthor{\bsnm{Jain}, \binits{P.}},
\bauthor{\bsnm{Shi}, \binits{Y.}},
\bauthor{\bsnm{Niranjan}, \binits{U.N.}}:
\bctitle{Tensor vs. matrix methods: Robust tensor decomposition under block
  sparse perturbations}.
In: \bbtitle{Artificial Intelligence and Statistics},
pp. \bfpage{268}--\blpage{276}
(\byear{2016}).
\bcomment{PMLR}
\end{bchapter}
\endbibitem

\bibitem{kajo2018svd}
\begin{barticle}
\bauthor{\bsnm{Kajo}, \binits{I.}},
\bauthor{\bsnm{Kamel}, \binits{N.}},
\bauthor{\bsnm{Ruichek}, \binits{Y.}},
\bauthor{\bsnm{Malik}, \binits{A.S.}}:
\batitle{Svd-based tensor-completion technique for background initialization}.
\bjtitle{IEEE Transactions on Image Processing}
\bvolume{27}(\bissue{6}),
\bfpage{3114}--\blpage{3126}
(\byear{2018})
\end{barticle}
\endbibitem

\bibitem{kajo2020tensor}
\begin{barticle}
\bauthor{\bsnm{Kajo}, \binits{I.}},
\bauthor{\bsnm{Kamel}, \binits{N.}},
\bauthor{\bsnm{Ruichek}, \binits{Y.}}:
\batitle{Tensor-based approach for background-foreground separation in maritime
  sequences}.
\bjtitle{IEEE Transactions on Intelligent Transportation Systems}
\bvolume{22}(\bissue{11}),
\bfpage{7115}--\blpage{7128}
(\byear{2020})
\end{barticle}
\endbibitem

\bibitem{kim2015compression}
\begin{botherref}
\oauthor{\bsnm{Kim}, \binits{Y.-D.}},
\oauthor{\bsnm{Park}, \binits{E.}},
\oauthor{\bsnm{Yoo}, \binits{S.}},
\oauthor{\bsnm{Choi}, \binits{T.}},
\oauthor{\bsnm{Yang}, \binits{L.}},
\oauthor{\bsnm{Shin}, \binits{D.}}:
Compression of deep convolutional neural networks for fast and low power mobile
  applications.
arXiv preprint arXiv:1511.06530
(2015)
\end{botherref}
\endbibitem

\bibitem{wang2016accelerating}
\begin{bchapter}
\bauthor{\bsnm{Wang}, \binits{P.}},
\bauthor{\bsnm{Cheng}, \binits{J.}}:
\bctitle{Accelerating convolutional neural networks for mobile applications}.
In: \bbtitle{Proceedings of the 24th ACM International Conference on
  Multimedia},
pp. \bfpage{541}--\blpage{545}
(\byear{2016})
\end{bchapter}
\endbibitem

\bibitem{novikov2015tensorizing}
\begin{botherref}
\oauthor{\bsnm{Novikov}, \binits{A.}},
\oauthor{\bsnm{Podoprikhin}, \binits{D.}},
\oauthor{\bsnm{Osokin}, \binits{A.}},
\oauthor{\bsnm{Vetrov}, \binits{D.P.}}:
Tensorizing neural networks.
Advances in neural information processing systems
\textbf{28}
(2015)
\end{botherref}
\endbibitem

\bibitem{hubara2017quantized}
\begin{barticle}
\bauthor{\bsnm{Hubara}, \binits{I.}},
\bauthor{\bsnm{Courbariaux}, \binits{M.}},
\bauthor{\bsnm{Soudry}, \binits{D.}},
\bauthor{\bsnm{El-Yaniv}, \binits{R.}},
\bauthor{\bsnm{Bengio}, \binits{Y.}}:
\batitle{Quantized neural networks: Training neural networks with low precision
  weights and activations}.
\bjtitle{The Journal of Machine Learning Research}
\bvolume{18}(\bissue{1}),
\bfpage{6869}--\blpage{6898}
(\byear{2017})
\end{barticle}
\endbibitem

\bibitem{hinton2015distilling}
\begin{botherref}
\oauthor{\bsnm{Hinton}, \binits{G.}},
\oauthor{\bsnm{Vinyals}, \binits{O.}},
\oauthor{\bsnm{Dean}, \binits{J.}}, et al.:
Distilling the knowledge in a neural network.
arXiv preprint arXiv:1503.02531
\textbf{2}(7)
(2015)
\end{botherref}
\endbibitem

\bibitem{romero2014fitnets}
\begin{botherref}
\oauthor{\bsnm{Romero}, \binits{A.}},
\oauthor{\bsnm{Ballas}, \binits{N.}},
\oauthor{\bsnm{Kahou}, \binits{S.E.}},
\oauthor{\bsnm{Chassang}, \binits{A.}},
\oauthor{\bsnm{Gatta}, \binits{C.}},
\oauthor{\bsnm{Bengio}, \binits{Y.}}:
Fitnets: Hints for thin deep nets.
arXiv preprint arXiv:1412.6550
(2014)
\end{botherref}
\endbibitem

\bibitem{polino2018model}
\begin{botherref}
\oauthor{\bsnm{Polino}, \binits{A.}},
\oauthor{\bsnm{Pascanu}, \binits{R.}},
\oauthor{\bsnm{Alistarh}, \binits{D.}}:
Model compression via distillation and quantization.
arXiv preprint arXiv:1802.05668
(2018)
\end{botherref}
\endbibitem

\bibitem{seddik2020lightweight}
\begin{bchapter}
\bauthor{\bsnm{Seddik}, \binits{M.E.A.}},
\bauthor{\bsnm{Essafi}, \binits{H.}},
\bauthor{\bsnm{Benzine}, \binits{A.}},
\bauthor{\bsnm{Tamaazousti}, \binits{M.}}:
\bctitle{Lightweight neural networks from pca \& lda based distilled dense
  neural networks}.
In: \bbtitle{2020 IEEE International Conference on Image Processing (ICIP)},
pp. \bfpage{3060}--\blpage{3064}
(\byear{2020}).
\bcomment{IEEE}
\end{bchapter}
\endbibitem

\bibitem{andriantsiory2021multi}
\begin{bchapter}
\bauthor{\bsnm{Andriantsiory}, \binits{D.F.}},
\bauthor{\bsnm{Geloun}, \binits{J.B.}},
\bauthor{\bsnm{Lebbah}, \binits{M.}}:
\bctitle{Multi-slice clustering for 3-order tensor}.
In: \bbtitle{2021 20th IEEE International Conference on Machine Learning and
  Applications (ICMLA)},
pp. \bfpage{173}--\blpage{178}
(\byear{2021}).
\bcomment{IEEE}
\end{bchapter}
\endbibitem

\bibitem{lebedev2014speeding}
\begin{botherref}
\oauthor{\bsnm{Lebedev}, \binits{V.}},
\oauthor{\bsnm{Ganin}, \binits{Y.}},
\oauthor{\bsnm{Rakhuba}, \binits{M.}},
\oauthor{\bsnm{Oseledets}, \binits{I.}},
\oauthor{\bsnm{Lempitsky}, \binits{V.}}:
Speeding-up convolutional neural networks using fine-tuned cp-decomposition.
arXiv preprint arXiv:1412.6553
(2014)
\end{botherref}
\endbibitem

\bibitem{liu2015sparse}
\begin{bchapter}
\bauthor{\bsnm{Liu}, \binits{B.}},
\bauthor{\bsnm{Wang}, \binits{M.}},
\bauthor{\bsnm{Foroosh}, \binits{H.}},
\bauthor{\bsnm{Tappen}, \binits{M.}},
\bauthor{\bsnm{Pensky}, \binits{M.}}:
\bctitle{Sparse convolutional neural networks}.
In: \bbtitle{Proceedings of the IEEE Conference on Computer Vision and Pattern
  Recognition},
pp. \bfpage{806}--\blpage{814}
(\byear{2015})
\end{bchapter}
\endbibitem

\bibitem{wen2016learning}
\begin{botherref}
\oauthor{\bsnm{Wen}, \binits{W.}},
\oauthor{\bsnm{Wu}, \binits{C.}},
\oauthor{\bsnm{Wang}, \binits{Y.}},
\oauthor{\bsnm{Chen}, \binits{Y.}},
\oauthor{\bsnm{Li}, \binits{H.}}:
Learning structured sparsity in deep neural networks.
Advances in neural information processing systems
\textbf{29}
(2016)
\end{botherref}
\endbibitem

\bibitem{allen2012sparse}
\begin{bchapter}
\bauthor{\bsnm{Allen}, \binits{G.}}:
\bctitle{Sparse higher-order principal components analysis}.
In: \bbtitle{Artificial Intelligence and Statistics},
pp. \bfpage{27}--\blpage{36}
(\byear{2012}).
\bcomment{PMLR}
\end{bchapter}
\endbibitem

\bibitem{vasilescu2004tensortextures}
\begin{bchapter}
\bauthor{\bsnm{Vasilescu}, \binits{M.A.O.}},
\bauthor{\bsnm{Terzopoulos}, \binits{D.}}:
\bctitle{Tensortextures: Multilinear image-based rendering}.
In: \bbtitle{ACM SIGGRAPH 2004 Papers},
pp. \bfpage{336}--\blpage{342}
(\byear{2004})
\end{bchapter}
\endbibitem

\bibitem{vlasic2006face}
\begin{bchapter}
\bauthor{\bsnm{Vlasic}, \binits{D.}},
\bauthor{\bsnm{Brand}, \binits{M.}},
\bauthor{\bsnm{Pfister}, \binits{H.}},
\bauthor{\bsnm{Popovic}, \binits{J.}}:
\bctitle{Face transfer with multilinear models}.
In: \bbtitle{ACM SIGGRAPH 2006 Courses},
p. \bfpage{24}
(\byear{2006})
\end{bchapter}
\endbibitem

\bibitem{liu2012denoising}
\begin{barticle}
\bauthor{\bsnm{Liu}, \binits{X.}},
\bauthor{\bsnm{Bourennane}, \binits{S.}},
\bauthor{\bsnm{Fossati}, \binits{C.}}:
\batitle{Denoising of hyperspectral images using the parafac model and
  statistical performance analysis}.
\bjtitle{IEEE Transactions on Geoscience and Remote Sensing}
\bvolume{50}(\bissue{10}),
\bfpage{3717}--\blpage{3724}
(\byear{2012})
\end{barticle}
\endbibitem

\bibitem{prevost2020hyperspectral}
\begin{barticle}
\bauthor{\bsnm{Pr{\'e}vost}, \binits{C.}},
\bauthor{\bsnm{Usevich}, \binits{K.}},
\bauthor{\bsnm{Comon}, \binits{P.}},
\bauthor{\bsnm{Brie}, \binits{D.}}:
\batitle{Hyperspectral super-resolution with coupled tucker approximation:
  Recoverability and svd-based algorithms}.
\bjtitle{IEEE Transactions on Signal Processing}
\bvolume{68},
\bfpage{931}--\blpage{946}
(\byear{2020})
\end{barticle}
\endbibitem

\bibitem{bourennane2010improvement}
\begin{barticle}
\bauthor{\bsnm{Bourennane}, \binits{S.}},
\bauthor{\bsnm{Fossati}, \binits{C.}},
\bauthor{\bsnm{Cailly}, \binits{A.}}:
\batitle{Improvement of classification for hyperspectral images based on tensor
  modeling}.
\bjtitle{IEEE Geoscience and Remote Sensing Letters}
\bvolume{7}(\bissue{4}),
\bfpage{801}--\blpage{805}
(\byear{2010})
\end{barticle}
\endbibitem

\bibitem{bourennane2010improvementB}
\begin{bchapter}
\bauthor{\bsnm{Bourennane}, \binits{S.}},
\bauthor{\bsnm{Fossati}, \binits{C.}},
\bauthor{\bsnm{Cailly}, \binits{A.}}:
\bctitle{Improvement of target detection based on tensorial modelling}.
In: \bbtitle{2010 18th European Signal Processing Conference},
pp. \bfpage{304}--\blpage{308}
(\byear{2010}).
\bcomment{IEEE}
\end{bchapter}
\endbibitem

\bibitem{dey2019tensor}
\begin{barticle}
\bauthor{\bsnm{Dey}, \binits{N.}},
\bauthor{\bsnm{Hong}, \binits{S.}},
\bauthor{\bsnm{Ach}, \binits{T.}},
\bauthor{\bsnm{Koutalos}, \binits{Y.}},
\bauthor{\bsnm{Curcio}, \binits{C.A.}},
\bauthor{\bsnm{Smith}, \binits{R.T.}},
\bauthor{\bsnm{Gerig}, \binits{G.}}:
\batitle{Tensor decomposition of hyperspectral images to study autofluorescence
  in age-related macular degeneration}.
\bjtitle{Medical image analysis}
\bvolume{56},
\bfpage{96}--\blpage{109}
(\byear{2019})
\end{barticle}
\endbibitem

\bibitem{decurninge2021tensor}
\begin{bchapter}
\bauthor{\bsnm{Decurninge}, \binits{A.}},
\bauthor{\bsnm{Land}, \binits{I.}},
\bauthor{\bsnm{Guillaud}, \binits{M.}}:
\bctitle{Tensor decomposition bounds for tbm-based massive access}.
In: \bbtitle{2021 IEEE 22nd International Workshop on Signal Processing
  Advances in Wireless Communications (SPAWC)},
pp. \bfpage{346}--\blpage{350}
(\byear{2021}).
\bcomment{IEEE}
\end{bchapter}
\endbibitem

\bibitem{henrion1994n}
\begin{barticle}
\bauthor{\bsnm{Henrion}, \binits{R.}}:
\batitle{N-way principal component analysis theory, algorithms and
  applications}.
\bjtitle{Chemometrics and intelligent laboratory systems}
\bvolume{25}(\bissue{1}),
\bfpage{1}--\blpage{23}
(\byear{1994})
\end{barticle}
\endbibitem

\bibitem{sun2020opportunities}
\begin{botherref}
\oauthor{\bsnm{Sun}, \binits{W.}},
\oauthor{\bsnm{Braatz}, \binits{R.D.}}:
Opportunities in tensorial data analytics for chemical and biological
  manufacturing processes.
Computers \& Chemical Engineering,
107099
(2020)
\end{botherref}
\endbibitem

\bibitem{ma2013three}
\begin{barticle}
\bauthor{\bsnm{Ma}, \binits{J.}}:
\batitle{Three-dimensional irregular seismic data reconstruction via low-rank
  matrix completion}.
\bjtitle{Geophysics}
\bvolume{78}(\bissue{5}),
\bfpage{181}--\blpage{192}
(\byear{2013})
\end{barticle}
\endbibitem

\bibitem{mahyari2016tensor}
\begin{barticle}
\bauthor{\bsnm{Mahyari}, \binits{A.G.}},
\bauthor{\bsnm{Zoltowski}, \binits{D.M.}},
\bauthor{\bsnm{Bernat}, \binits{E.M.}},
\bauthor{\bsnm{Aviyente}, \binits{S.}}:
\batitle{A tensor decomposition-based approach for detecting dynamic network
  states from eeg}.
\bjtitle{IEEE Transactions on Biomedical Engineering}
\bvolume{64}(\bissue{1}),
\bfpage{225}--\blpage{237}
(\byear{2016})
\end{barticle}
\endbibitem

\bibitem{nion2010tensor}
\begin{barticle}
\bauthor{\bsnm{Nion}, \binits{D.}},
\bauthor{\bsnm{Sidiropoulos}, \binits{N.D.}}:
\batitle{Tensor algebra and multidimensional harmonic retrieval in signal
  processing for mimo radar}.
\bjtitle{IEEE Transactions on Signal Processing}
\bvolume{58}(\bissue{11}),
\bfpage{5693}--\blpage{5705}
(\byear{2010})
\end{barticle}
\endbibitem

\bibitem{tichai2019tensor}
\begin{barticle}
\bauthor{\bsnm{Tichai}, \binits{A.}},
\bauthor{\bsnm{Schutski}, \binits{R.}},
\bauthor{\bsnm{Scuseria}, \binits{G.E.}},
\bauthor{\bsnm{Duguet}, \binits{T.}}:
\batitle{Tensor-decomposition techniques for ab initio nuclear structure
  calculations: From chiral nuclear potentials to ground-state energies}.
\bjtitle{Physical Review C}
\bvolume{99}(\bissue{3}),
\bfpage{034320}
(\byear{2019})
\end{barticle}
\endbibitem

\end{thebibliography}


\end{document}